\documentclass{aa} % Standard layout

\usepackage[varg]{txfonts} 
\usepackage{xcolor}
\usepackage{natbib,twoopt}
\usepackage[breaklinks=true]{hyperref} %% to avoid \citeads line fills
\bibpunct{(}{)}{;}{a}{}{,} %% natbib format for A&A and ApJ

\makeatletter
  \newcommandtwoopt{\citeads}[3][][]{\href{http://adsabs.harvard.edu/abs/#3}%
    {\def\hyper@linkstart##1##2{}%
     \let\hyper@linkend\@empty\citealp[#1][#2]{#3}}}
  \newcommandtwoopt{\citepads}[3][][]{\href{http://adsabs.harvard.edu/abs/#3}%
    {\def\hyper@linkstart##1##2{}%
     \let\hyper@linkend\@empty\citep[#1][#2]{#3}}}
  \newcommandtwoopt{\citetads}[3][][]{\href{http://adsabs.harvard.edu/abs/#3}%
    {\def\hyper@linkstart##1##2{}%
     \let\hyper@linkend\@empty\citet[#1][#2]{#3}}}
  \newcommandtwoopt{\citeyearads}[3][][]%
    {\href{http://adsabs.harvard.edu/abs/#3}
    {\def\hyper@linkstart##1##2{}%
     \let\hyper@linkend\@empty\citeyear[#1][#2]{#3}}}
    \renewcommand*\aa@pageof{, page \thepage{} of \pageref*{LastPage}}
\makeatother

\newcommand{\Vrad}{$V_{\rm Rad}$}
\newcommand{\alfa}{$\alpha$}
\newcommand{\alfaFe}{[$\alpha$/Fe]}
\newcommand{\meta}{[M/H]}
\newcommand{\FeH}{[Fe/H]}
\newcommand{\T}{$T_{\rm eff}$}
\newcommand{\g}{log($g$)}
\newcommand{\gunits}{cm/s$^2$}
\newcommand{\SH}{[S/H]}
\newcommand{\SFe}{[S/Fe]}
\newcommand{\SM}{[S/M]}

\begin{document}
        
        \title{The AMBRE Project: Origin and evolution of sulfur\\ in the Milky 
        Way\thanks{Tables 5 and 6 are available in electronic form
at the CDS via anonymous ftp to cdsarc.u-strasbg.fr (130.79.128.5)
or via http://cdsweb.u-strasbg.fr/cgi-bin/qcat?J/A+A/}}
        %\subtitle{HARPS,UVES,FEROS}

%\institute{U}

        \author{
                J. Perdigon\thanks{Send offprint requests to Patrick de Laverny}\inst{1}
                \and 
                P. de Laverny\inst{1}
                \and 
                A. Recio-Blanco \inst{1}  
                \and
                E. Fernandez-Alvar\inst{1} 
                \and 
                P. Santos-Peral\inst{1} 
                \and 
                G. Kordopatis\inst{1} 
                \and 
                M.A. \'Alvarez\inst{2} }
        
        \institute{ 
        Universit\'e C\^{o}te d'Azur, Observatoire de la C\^{o}te d'Azur, CNRS, Laboratoire Lagrange, Bd de l'Observatoire, CS 34229, 06304 Nice cedex 4, France.
        \and
        %Universidade da Coruña, Provincia Autonoma de Galicia, Kingdom of Spain!!!!!!!
        CIGUS -CITIC- Department of Computer Science and Information Technologies, University of A Coru\~na, Spain
        }
        
        %\thanks{<text of footnote>}
        
        \date{Received <date> / Accepted <date>}
        
        \abstract{Sulfur is a volatile chemical element that plays an important role in tracing the
        chemical evolution of the Milky Way and external galaxies. However, its nucleosynthesis origin and abundance 
        variations in the Galaxy are still unclear because the 
        number of available stellar sulfur abundance measurements is currently rather small.}
        {The goal of the present article is to accurately and precisely study the sulfur content of large number of stars located in the solar neighbourhood.}
        {We use the parametrisation  of thousands of high-resolution  stellar spectra provided by the AMBRE Project, and combine it with the automated abundance determination GAUGUIN to derive  local thermodynamic equilibrium (LTE) sulfur abundances for 1855 slow-rotating FGK-type stars.  
        This is the largest and most precise catalogue of sulfur abundances  published to date. It covers a
        metallicity domain as high as $\sim$2.5~dex starting at \meta $\sim$-2.0~dex.}
        {We find that the sulfur-to-iron abundances ratio is compatible with a plateau-like 
        distribution in the metal-poor regime, 
        and then starts to decrease continuously at \meta $\sim$-1.0~dex. This decrease continues
        towards negative values for supersolar metallicity stars as recently reported for magnesium and as predicted by Galactic chemical 
        evolution models. Moreover, sulfur-rich stars having metallicities in the range [-1.0,-0.5] have very different kinematical and orbital properties 
        with respect to more metal-rich and sulfur-poor ones. Two disc components, associated with the thin and thick discs,
        are thus seen independently in kinematics and sulfur abundances. The sulfur radial gradients
        in the Galactic discs have also been estimated. Finally, the enrichment in sulfur with respect
        to iron is nicely correlated with stellar ages: older metal-poor stars have higher [S/M] ratios than
        younger metal-rich ones.}
        {This work has confirmed that sulfur is an \alfa-element that could be considered
        to explore the Galactic populations properties. For the first time, a chemo-dynamical study
from the sulfur  abundance point of view, as a stand-alone chemical element, is   performed.}
        
        
        \maketitle
        
        \section{Introduction}

        Sulfur is a chemical species of particular importance in the context of stellar nucleosynthesis and the chemical evolution of galaxies. It is a volatile element and, as a consequence, it is not blocked into the dust grains of the
interstellar medium (ISM). It is therefore a good tracer of the chemical evolution of galaxies, in particular
at large redshifts \citep{Savage96}. Moreover, from the stellar nucleosynthesis point of view, sulfur is classified as an \alfa-element (e.g. oxygen, magnesium, titanium, ...).
It is indeed produced via \alfa-capture in the inner layers 
of massive stars \citep[see e.g.][]{Woosley95, 2013ARA&A..51..457N}. These chemical species are then  released into the ISM mostly through  Type II supernovae on a relatively short timescale. 
It is therefore believed that the abundance of all these $\alpha$-elements approximately follows the same behaviour during the Galactic chemical evolution.
On the other hand, although also partly produced in massive stars, iron is mostly produced and expelled into the ISM by Type Ia supernovae  on a much longer timescale. Thus, we expect that the \alfaFe \ ratio (and thus \SFe ) remains almost constant with respect to iron content in the metal-poor regime (i.e. as a plateau),
corresponding to the epoch before the ignition of the first Type Ia supernovae. Then, this ratio is expected to decrease afterwards as soon as the amount of released iron increases with time. 
Adopting such production sites, Galactic chemical evolution models quite successfully reproduce the main observed behaviour of sulfur and other \alfa-elements as a function of the metallicity \citep[among other recent studies, see][]{Nikos18, 
Valeria17, Palla20}, confirming that the major production site of S is indeed Type II supernovae.
        
        On the observational side, a few  studies have reported sulfur abundance enhancements in the metal-poor regime. \citet{Francois88} was the first to suspect the \alfa-like behaviour of sulfur for \FeH~$<$~-1~dex with a plateau-like structure, and this was then confirmed by later studies \citep[see e.g.][]{Nils04, Nissen04, Nissen07, Spite11, Kacharov15}. These works invalidated the suggestion of \cite{Garik01} who found a steady increase in \SFe \  with decreasing metallicities, leading to some suspected extreme sulfur-rich stars (\SFe $\approx$ 0.8 dex at \FeH < -2.0 dex). The plateau of sulfur in the metal-poor regime nonetheless appears with a large dispersion mainly caused by the difficulty of analysing weak lines at these low metallicities. 
        At higher but still subsolar metallicities (-1.0~dex $\la$ \FeH $\la$ 0.0~dex), and as expected for any  \alfa -element, the decline of [S/Fe] with metallicity in the Galactic disc is clearly observed and interpreted as the release of iron from Type Ia SNe at [Fe/H] $\la$ -1~dex \citep[see e.g.][]{2002A&A...390..225C, Caffau05, Nils06, Nils13}. This therefore supports the idea of a common nucleosynthetic origin for sulfur and other \alfa -species.
        
It is interesting to note that all these studies rely on rather small samples of stars (typically a few tens). However, in order to be able to study possible different sulfur content in the various Galactic populations, large statistics is necessary. This changed recently thanks to a few studies.
First, \cite{Luck15} reported S{\sc i} abundances of $\sim$ 1,100 G-K giants but with rather large uncertainties and dispersions in \SFe. Moreover, a strong temperature dependency, probably caused by blending from unknown lines, caused blurring in the main picture. Then, \cite{Takeda16} derived sulfur abundances for up to $\sim 400$ dwarfs and giants, confirming the decrease in  \SFe \ with metallicity. Later on, within the $Gaia$-ESO Survey, \cite{GES17} managed to derive sulfur abundances for a sample of 1,301 Galactic stars, including stars in open and globular clusters, but with only half a dozen  stars 
below \meta $\sim$ -1.0~dex.  Although the global behaviour is again partially blurred by rather large
measurement dispersions and temperature-dependent derived abundances, this study seems to confirm the \alfa-like behaviour 
of sulfur at subsolar metallicities (-1 $\la$ \meta $\la$0~dex). Finally, \citet[hereafter CS20]{Porto20} recently presented a precise analysis of sulfur in 719 stars of the solar neighbourhood having metallicities
higher than -1.0~dex. This work clearly showed that sulfur behaves like a typical \alfa-element in the thin and thick discs,
in rather good agreement with the literature models of Galactic sulfur evolution \citep{Romano10, Koba11, Nikos18}.

        In the present study we  profit from the spectra parametrised within the AMBRE Project \citetads{AMBRE13} and provided by the ESO archives of the HARPS, FEROS, and UVES spectrographs. We estimated precise and homogeneous sulfur abundances for a catalogue containing an unprecedented number of stars, having metallicities varying from $\sim$-2.0~dex to $\sim$+0.7dex. This allowed us to depict a global and homogeneous view of sulfur in the main Galactic components.
        The paper is structured as follows. We present in Sect.~\ref{method} the method developed for automatically deriving the sulfur abundances. 
        The AMBRE-sulfur catalogue of almost 1,855 stars is then presented in Sect.~\ref{AMBRE_sulfur_catalogue}. In Sect.~\ref{Discussion}, we discuss the sulfur behaviour in the
        solar vicinity in terms of abundance variations, stellar kinematics and ages, and
        Galactic radial gradients. We finally summarise our results in Sect.~\ref{Conclusions}.

        \section{Derivation of the AMBRE sulfur abundances}\label{method}
        
        This study has been carried out in the framework of the AMBRE Project \citep{AMBRE13}, whose first aim was to derive the main atmospheric parameters (effective temperature \T , surface gravity \g , mean metallicity \meta, and enrichment in \alfa -elements with respect to iron  \alfaFe) of ESO archived spectra. This is performed thanks to the MATISSE algorithm \citep{Recio06}
        trained with a specific grid of synthetic spectra \citep{AMBRE12}. For the present analysis, we also use  other AMBRE data products  \citep[for a detailed description, see][]{Worley12} as the stellar radial velocity (\Vrad), the signal-to-noise ratio (S/N),
	the full width at half maximum (FWHM) of the cross-correlation function (CCF) used to estimate \Vrad \ (i.e. an estimate of the typical width of the lines, therefore including the effects of the rotational velocity), and a quality flag of the stellar parametrisation (based on the computation of a $\chi ^2$ between the observed and reconstructed spectra at the derived stellar parameters). 
        
        From these parametrised AMBRE spectra the sulfur abundances were then derived thanks to GAUGUIN, an optimisation  method coupling a precomputed grid of synthetic spectra (see Sect. 2.2. for a description of this grid) and a Gauss-Newton algorithm.     GAUGUIN was originally developed in the framework of the $Gaia$/RVS analysis within the $Gaia$/DPAC for the estimation of the stellar atmospheric parameters: for the mathematical basis, see \cite{GAUGUIN10}; and then, the first applications in 
	\cite{GAUGUIN12} and \cite{Recio16}. A natural and simple extension of GAUGUIN's applicability to the derivation of stellar chemical abundances was then initiated within the context of the $Gaia$/RVS 
\cite[DPAC/Apsis pipeline,][]{Apsis13},
the AMBRE Project and the $Gaia$-ESO Survey. We first published a detailed description of the application of GAUGUIN for the derivation of chemical abundances within the AMBRE context in \citet{GAUGUIN16}. We also refer to Sect.\ref{dataset} for its specific application to sulfur abundances.
        
        \subsection{Adopted line list and computation of the grid of reference synthetic spectra}\label{linelist}
        
        For the selection of our analysed sulfur lines, we refer to the works of \cite{Caffau05}, \citetads{GES17}, and \citetads{Takeda16}
        who studied  multiplets 1, 6, and 8 (respectively at $\approx$~922, 869, and 675~nm).  Taking into account (i) the wavelength ranges covered by our
        observed spectra (see Subsect.~\ref{dataset}), (ii) the strength of these S{\sc i} 
        multiplets in FGK star spectra, (iii) the existence of almost blend-free spectral 
        ranges, and (iv) the expected weak non-local thermodynamic equilibrium (NLTE) effects  (see below), we selected the lines of multiplet~8 for the present analysis. 
        It is  known that these  lines are almost unaffected by NLTE effects since they are formed
        in deep atmospheric layers \citep{Korotin09}. 
	For example, \citet{Takeda16} and \citet{Korotin17}
	have shown that NLTE departures should always be smaller than 0.1~dex for our sample stars.

        For the present analysis of the multiplet~8 lines the atomic data of \citet{Wiese69} were adopted and 
        are reported in Table~\ref{table:linelist}. 
        We note that we have found   a small systematic bias in the abundances
        derived from the 674.8~nm line with respect to the two others. This could reflect some
        possible small uncertainties in the atomic data adopted for the three components of this line (see Sect.~2.3.3).
        The proposed correction for the $\log{gf}$ (+0.08~dex) of these 674.8~nm transitions would   agree closely with the line
        data of \citet{Biemont93}.
        % the work of the $Gaia$-ESO Survey line-list group \citep{Heiter20}, except for the triplet around 675.7~nm \ for which the line data from \citetads{Caffau05} leads to a better fit of the Solar spectrum and, has thus been adopted in the present study. 
        
    \begin{table}[!t]
            \caption{Adopted multiplet 8 sulfur line data \citep{Wiese69}.}
        % title of Table
        \label{table:linelist}
        % is used to refer this table in the text
        \centering
        % used for centering table
        \begin{tabular}{l l l}
            % centered columns (4 columns)
            \hline
            % inserts double horizontal lines
            $line$~(nm) & $\chi_{e}$~(eV) &  $\log{gf}$ \\
            % table heading
            \hline  
                674.3440 & 7.866 & -1.27 \\
                674.3531 & 7.866 & -0.92 \\
                674.3640 & 7.866 & -1.03 \\                     
                674.8573 & 7.868 & -1.39\tablefootmark{*} \\
                674.8682 & 7.868 & -0.80\tablefootmark{*} \\
                674.8837 & 7.868 & -0.60\tablefootmark{*} \\
            %675.6851\tablefootmark{*} & 7.870\tablefootmark{*} & -1.76\tablefootmark{*} \\
            %675.7007\tablefootmark{*} & 7.870\tablefootmark{*} & -0.90\tablefootmark{*} \\
            %675.7171\tablefootmark{*} & 7.870\tablefootmark{*} & -0.31\tablefootmark{*} \\
            675.6851 & 7.870 & -1.76 \\
            675.7007 & 7.870 & -0.90 \\
            675.7171 & 7.870 & -0.31 \\
            \hline
        \end{tabular}
            %\tablefoot{\tablefoottext{*}{From \citetads{Wiese69}.}}
            \tablefoot{\tablefoottext{*}{Probably underestimated, see text.}}
    \end{table}
    
    From these adopted S{\sc i} lines completed by all the molecular and atomic linelists of \citet{Heiter20}, we computed a grid of synthetic spectra around the selected S{\sc i} lines thanks to the TURBOSPECTRUM code \citep{TURBO} and the MARCS model atmospheres \citep{MARCS} under the LTE, 1D, and hydrostatic assumptions, adopting the \cite{Grevesse07} solar chemical composition. For this specific sulfur grid, we followed a similar, but slightly updated, procedure as in \cite{AMBRE12}. The ranges of the atmospheric parameters are 4,000 $\le$ \T $\le$ 8,000~K (in steps of 250~K), +0.0 $\le \g \le$ +5.5 (in steps of 0.5, $g$ being in \gunits), and -5.0 $\le$ \meta $\le$ +1.0~dex (with steps of 0.5~dex for metallicities smaller than -1.0~dex and 0.25~dex above);  up to 13 values of \alfaFe \ were considered for each value of \meta\ depending on the availability of the MARCS models (with steps of 0.1~dex). Therefore, 25,961 MARCS model atmospheres with an \alfaFe \ enhancement consistent with that adopted when computing the synthetic spectra were considered. Then, for each combination of these four atmospheric parameters, we computed spectra by varying sulfur abundances between -3.0 $\le \SH \le$ +2.0~dex with a step of 0.2~dex (i.e. 26 different values of \SH). The adopted sulfur solar abundance is $A_{\rm S}$=7.12 \citep{Scott}. 
     For these computations, we also adopted a micro-turbulence velocity that varies 
     with \T, \g , and \meta \ following the prescription of the $Gaia$-ESO Survey (version 2 of the GES empirical relation based on microturbulence velocity  determinations from literature samples; Bergemann et al., in preparation). 
     The adopted  micro-turbulence velocities vary from 0.6 to 4.8~km/s, depending on the stellar parameters.
     We also recall that no stellar rotation is considered when computing the grid spectra. Finally, the reference AMBRE sulfur grid consists in %25,961$\times$ 11 sulfur abundances, i.e. a total of 
about 675,000 spectra, covering the wavelength range from 672 to 677~nm  with a wavelength step of 0.0005~nm.
    
    \subsection{Chemical analysis of  AMBRE: HARPS, UVES, FEROS spectra}
    \label{dataset} 
   
    Our data consists of a collection of about 100,000 ESO archived spectra from the FEROS, HARPS, and UVES spectrographs, already parametrised within the AMBRE Project (see  \cite{Worley12},  \cite{dePascale14} and \cite{Worley16}, respectively). The number of selected spectra having a S/N higher  than 20 and a quality flag for their AMBRE parametrisation equal to 0 or 1 (i.e. good or very good parametrisation; see the above AMBRE papers), together with their atmospheric parameter ranges and their mean signal-to-noise ratio is summarised in Table~\ref{table:sum_data}. We note  that two UVES setups cover the selected S{\sc i} multiplet~8 lines, and are thus identified as two separate spectrographs.
    
    In order to be analysed with our GAUGUIN pipeline in a homogeneous way, the observed spectra were first   corrected by their radial velocity. Then, the spectral resolutions of the HARPS and FEROS spectra were   degraded to the UVES value ($R \sim 40,000). $ We adopted a sampling wavelength step of 0.005~nm for the whole dataset in order to fulfil the Nyquist-Shannon criteria. The reference grid spectra was   convolved and re-sampled accordingly.
    
    Then, the prepared 99,271 spectra were ingested into our chemical analysis pipeline.
    The spectra are first automatically normalised by comparing a synthetic and an observed spectrum over
a  $\sim$7~nm  domain, centred on each sulfur line. Then, the normalisation is refined over a $\sim$0.4~nm interval around each line \citep[for more details on the normalisation procedure, see][]{Pablo20}.
    For each of these spectra, the three components of  the S{\sc i} multiplet 8 in Table~\ref{table:linelist} are then analysed independently by comparing the observed and reference grid line profiles over domains covering 0.06nm, 0.05nm, and 0.08nm, centred  at the S{\sc i}~674.3, 674.8, and 675.7~nm lines, respectively.
    We  end up with a catalogue of about 300,000 sulfur measurements, including several non-detection and upper limit measurements.
    
    \begin{table*}
        \caption{Summary of the parameter ranges covered by our selected AMBRE spectra having a signal-to-noise ratio higher than 20.}
        % title of Table
        \label{table:sum_data}
        % is used to refer this table in the text
        \centering
        % used for centering table
        \begin{tabular}{l c c c c c c c}
            % centered columns (7 columns)
            \hline
            % inserts double horizontal lines
            Spectrograph & Number of spectra & \T~(K) & \g ~($g$ in cm/s$^2$) & \meta~(dex) & \alfaFe~(dex) & <S/N> & $\sigma$(S/N) \\
            \hline
            % table heading
            HARPS & 88,178 & [4015, 7620] & [1.02, 4.95] & [-3.43, 0.61] & [-0.39, 0.59] & 74 & 29  \\
            FEROS & 5,821 & [4000, 7623] & [1.00, 4.99] & [-3.49, 0.95] & [-0.38, 0.65] & 98 & 52 \\
            UVES / Red580 & 3,533 & [3668, 7575] & [1.10, 5.00] & [-3.47, 0.72] & [-0.33, 0.50] & 192 & 89 \\
            UVES / Red860 & 1,739 & [3584, 7787] & [0.00, 4.82] & [-3.49, 0.75] & [-0.39, 0.79] & 157 & 51 \\
            \hline
        \end{tabular}
%        \tablefoot{\tablefoottext{1}{with $SNR\geq20$}. \tablefoottext{2}{Atmospheric parameters ranges covered by our data}}
    \end{table*}
    
    %Ce qui sort de GAUGUIN et comment on construit le cat. final
    
        \subsection{Construction of the AMBRE-sulfur catalogue}
        
        The AMBRE-sulfur 
	catalogue that is presented and discussed in the next sections was   built as follows. We first note that our dataset contains, for some stars, a large number of repeated spectra (hereafter called `{\it repeats}'). We thus describe below how we derived sulfur abundances from the analysis of several {\it repeats} of the same star from which up to three distinct lines can be measured.
        
        \subsubsection{Cross-match with the $Gaia$ DR2 and adopted ID}
        
        The AMBRE spectra were   collected with quite different instruments. The available spectra may therefore contain heterogeneous  names of targets and accuracy of   coordinates. The first step was thus to identify the corresponding observed stars, and in particular the identification of the spectra belonging to the same star. For this purpose we made use of the $Gaia$ DR2 catalogue \citep{gai18} and adopted, when found, the $Gaia$ DR2 ID. 
We refer to a forthcoming article for the detailed presentation of this cross-match between the $Gaia$/DR2 and the AMBRE catalogues. 
Briefly, this cross-match was   performed using the 
 stellar spectra coordinates and different checks between the derived AMBRE atmospheric parameters, \T \ estimated from ground based photometry
\citep[2MASS and APASS,][]{2MASS, APASS}, and 
$Gaia$ data (e.g. G-magnitude, $B_P - R_P$ colours, \T, radial velocities,  but for more details, see also Santos-Peral, 2021, submitted). 
We were able to identify 5,076 distinct stars from the 99,271 AMBRE spectra. There is also a significant fraction ($\sim 20\%$) of our spectra for which no $Gaia$ DR2 ID were found. Several of these spectra actually correspond to bright stars, absent from the $Gaia$/DR2 catalogue. In this case, we simply looked 
in Simbad for stars having  coordinates similar to  the AMBRE coordinates within a radius of $10"$ and consistent parametrisation. We then adopted for them the corresponding name in the {\it Henry Draper} catalogue as an ID, adding about 200 more stars into the initial sample.

        Finally, among the identified stars, a significant part has several $repeats$: about $\sim$ 20\% of the stars have more than ten repeats, and more than ten stars have more than 1,000 associated spectra. Such a large number of 
$repeats$ allowed us to derive sulfur abundances with very low internal uncertainties (see below).
        
        \subsubsection{Selection of the best analysed spectra}\label{outliers}
        
        For a given star, we decided to only keep  the {\it repeats} with a  very consistent set of stellar parameters among each other. We therefore rejected the spectra that depart too much from the median values of all repeats. The threshold was   arbitrarily chosen to be 1~km/s, 100~K, 0.5, 0.10~dex, and 0.05~dex for Vrad,
        \T, \g, \meta, \ and \alfaFe , respectively. A unique threshold can be adopted for
        any spectrum since we recall that the AMBRE parametrisation
        was   performed at a constant spectral resolution for the three considered ESO 
        spectrographs. We also note that such a rejection procedure could help
        to reject possible spectroscopic binaries for which their stellar parameters may seem to vary
        between different epochs of observation.
        
        On the other hand, the GAUGUIN pipeline was not able in some cases to derive a useful 
        sulfur abundance. 
        This directly results from either a low spectrum quality (photon noise and/or cosmic rays) or from some atmospheric parameter limitations in the synthetic grid. For example, in the present analysis the reference grid is only valid  for
        low-rotation stars. However, the AMBRE parametrisation provides an indication of the line broadening (and hence of the rotational velocity, among other broadening mechanisms) thanks to the FWHM of the cross-correlation functions derived during the radial velocity measurement. Therefore, in order to reject stars whose   rotational and/or macroturbulent velocities were possibly too high, we systematically rejected all the spectra with a $FWHM_{\rm CCF}>15$~km/s for the three spectrographs, in agreement with previous estimates \citepads[see e.g.]{GAUGUIN16}. This value corresponds to rotational velocities typically lower than $\sim$10-15~km/s
        (depending on the stellar types) at the working spectral resolution.
        
        Moreover, for each analysed spectrum and detected sulfur lines, we also computed the lowest abundance (\textit{upper limit}) that could be estimated from their atmospheric parameters and signal-to-noise ratio. We systematically rejected all the derived abundances that were lower than or too close to twice this \textit{upper limit}. 
        
        Finally, all these different criteria led to the rejection of about $2/3$ of the measured sulfur lines, leading to about 100,000 useful measurements of one of the three 
        multiplet 8 sulfur lines.
        
        \subsubsection{Mean sulfur abundances and atmospheric parameters per star} \label{derivation_stellar_ab}
        We recall that our final working sample  consists of 5,275 distinct stars, of which the vast majority possess {\it repeat} spectra;  each of them could have up to three measured lines. 
Our method for determining the sulfur abundance of a given star consists of two steps: (i)   determining 
separately the mean abundance of the three sulfur lines for the available $\textit{repeats}$ and  (ii) estimating the final sulfur abundance of each star by averaging   up to three available mean individual abundances from the previous step. In both stages the same averaging method is  employed, following the work of \citetads[][hereafter A15]{2015A&A...583A..94A}. These authors investigated different methods of combining abundances extracted from different lines of a given element, and we adopted their weighted mean  (\textit{WM}) procedure to estimate our sulfur abundances. We briefly describe below the adopted methodology, and refer to A15 for more details.
        
%For a set of $N$ sulfur abundances $\left\lbrace[S/H]_i\right\rbrace$, the adopted $WM$ is:
        For a set of $N$ sulfur abundances [S/H]$_i$ the adopted $WM$ is
        
        \begin{equation}
                {\rm [S/H]} = \frac{\sum_{i=1}^{N}W_i ~ [S/H]_i}{\sum_{i=1}^{N}W_i} \label{eq:WM}
        ,\end{equation}
        
        \noindent with $W_i$ being $weights$ defined below. As shown by A15, this  \textit{WM} procedure has the advantage of successfully removing the effect of outliers on the final abundance, without using any 
ad hoc sigma-clipping procedure. The definition of the weights is based on the distance (in terms of standard deviation, $std$) between the available abundances and their median ($med$):
        
        \begin{equation}
        W_i = \frac{1}{dist_i} ~~~;~~~ dist_i = \frac{[S/H]_i - med\left\lbrace[S/H]\right\rbrace}{std\left\lbrace[S/H]\right\rbrace}\label{eq:W}
    .\end{equation}
    
    \noindent In practice, if more than half of the \textit{repeats} have the same abundance
    value, and if the abundance distribution is Gaussian, the associated median of the absolute
deviation (MAD), and hence the standard deviation, 
could be equal to zero, resulting in an infinite weight. To circumvent this effect we decided, as suggested by A15, to bin the $dist_i$ in boxes having a width equal to 0.5. For example, all the abundances with $dist_i \le 0.5$ will have the same weight ($1/0.5=2$ in this example). 
    
    We first applied this procedure independently for the three sulfur lines (step~(i) above), hence providing their $WM$ abundance. 
    The final sulfur abundance (step~(ii) above) estimated when more than one sulfur line was measured is also obtained with the same procedure. However, we note  that if only two S{\sc i} lines are available, we provide their $WM$ final
    abundance only if an abundance of the strongest S{\sc i}~675.7~nm component (and easiest line
to analyse) is available.
     If only one line has been measured for a given star, we accept its \SH  \ only if it comes from the 675.7~nm component.
    Finally, each line abundance is associated  with an error $\Delta \SH$, proportional to the $MAD$ amongst the \textit{repeats} of the line. We adopted $\Delta \SH = 1.483 \times MAD(\left\lbrace \SH_i\right\rbrace)$, 
    i.e. the scaled MAD (corresponding to a $1\sigma$ threshold for a Gaussian distribution).    

    We note that the atmospheric parameters (including the S/N) associated with a
    given star have been averaged by adopting  the same $WM$ procedure. The
mean dispersions associated with  these
means are equal to 6~K, 0.01, 0.006~dex, and 0.004~dex for \T , \g , \meta, \alfaFe, respectively.
    We also specify that the final sulfur abundance mean was   performed after   the 674.8~nm line was corrected for a systematic bias of -0.08~dex in \SH \ that could be associated with the non-calibrated line data
    (see Sect.~2.1). We validated our whole analysis procedure (including the line atomic data) by estimating the solar sulfur abundance derived thanks to a very high-S/N HARPS 
spectrum of Vesta and the solar FTS spectra of \cite{Wallace11}. 
For both spectra degraded at $R \sim 40,000$, we obtained \SH = -0.04~dex after correcting 
the 674.8~nm line. 
Such a small bias was also seen in other reference stars (assuming [S/\alfa]=0 for these stars).
We therefore calibrated all our final abundances by this small amount of -0.04~dex in order to be consistent
with the previously adopted solar sulfur abundances of \citet{Scott}.
%  Tests calibation
% 1 - Spectre Soleil /Vesta (5772/4.44/0.0/0.0)
%--> Vesta: Raie1/2/3=-0.07/+0.04-0.08?/-0.03 => [S/Fe]~-0.04
%--> SunHinkle: Raie1/2/3=-0.1/+0.02-0.08?/-0.06 => [S/Fe]~-0.05
%
%--> Arcturus/AMBRE+Hinkle: spectres jetés. Pb normalisation??
%
%--> Procyon: pb largeur raies? Vmacro pas pris en compte bien que R=40,000
%      Raie1/2/3=-0.06/-0.01-0.08?/-0.03 => [S/Fe]~-0.05

    Finally, the uncertainties were then estimated using again this $WM$ procedure, adopting the same weights as those adopted for the final abundance. If no $repeats$ are available  for some stars, the final errors were estimated from the dispersion between the abundance of the accepted individual lines. Therefore, no dispersions
are reported for stars having only one available spectra in which only one line has been measured (the S{\sc i}~675.7~nm component).
    
        \subsubsection{Uncertainties associated with the derived sulfur abundances}
        
        Several sources of uncertainties can affect an abundance determination. Moreover, looking at their different significances can help to flag the reported abundances and clean the sample for future scientific exploitation.
        %\footnote{We thank the present director of the Côte d'Azur Observatory for its numerous initiatives to ask us to focus less and less on the scientific exploitation of our abundance catalogues.}
        
	Firstly, we  estimated the sensitivity of the derived sulfur abundances due to possible uncertainties on the stellar atmospheric parameters. Table~\ref{Tab_error2}  presents the mean variations of \SH \ when considering 
        typical AMBRE uncertainties ($external$ errors estimated by comparison with external catalogues)
on \T , \g , and \meta \ for the different stellar types contained in our sample. 
We note that typical uncertainties on the microturbulence velocity ($\pm$1~km/s) and \alfaFe \
($\pm$0.1~dex) have no measurable effect on \SH. 
        It can be seen that the sulfur abundances are mostly
        sensitive to the \T \ uncertainties, and this effect is stronger for the coolest stars in which sulfur lines are weaker. The reported total uncertainties on the sulfur abundances have been estimated by summing quadratically the different contributions.

                   \begin{table*}[!t]
        \caption{Sensitivities of the sulfur abundances (in dex) caused by typical uncertainties on the stellar atmospheric parameters.}
        \label{Tab_error2}
        \centering
        \begin{tabular}{l c c c c}
            \hline
        &   Cool giant & Cool dwarf & Solar-type & Hot dwarf \\
        &   \T$\sim$4500~K & \T$\sim$5000~K & \T$\sim$5800~K & \T$\sim$6500~K \\ 
            \hline  
         $\Delta$ \T  \ = $\pm$120K & $\pm$0.20 & $\pm$0.20 & $\pm$0.05 & $\pm$0.03\\
         $\Delta$ \g  \ = $\pm$0.25dex & $\pm$0.07 & $\pm$0.05 & $\pm$0.05 & $\pm$0.04   \\
       $\Delta$ \meta \ = $\pm$0.10dex & $\pm$0.01 & $\pm$0.02 & $\pm$0.02 & $\pm$0.02   \\
       Total (quadratic sum) & $\pm$0.21 & $\pm$0.21 & $\pm$0.07 & $\pm$0.05 \\ 
                   \hline
        \end{tabular}
    \end{table*}

        Then, since we were   able to analyse several spectra of the same stars ($repeats$) spanning a wide range of S/N, we   checked the robustness of our automatic procedure for deriving sulfur abundances. The AMBRE large sample of $repeat$ spectra  allowed us to precisely quantify our typical $internal$ uncertainties that could result from several effects (e.g. continuum normalisation, spectra quality, radial velocity correction, differences in the atmospheric parameters). We list in Table~\ref{Tab_error1} the $internal$ errors for different stellar types. 
        For a given line and a given stellar type it can be seen that  the sulfur abundance 
        scatter from spectra to spectra is already very small at rather low S/N,
        and can even be smaller for higher S/N.
        The robustness of our automatic procedure is therefore confirmed.
        Moreover, we note that our final sulfur abundances are computed by averaging  the $repeated$ measurements and up to three sulfur lines, when available.  Consequently, the relative uncertainties from star to star in the AMBRE-sulfur catalogue are therefore even smaller than the  values given in Table~\ref{Tab_error1}. In any case, they can be neglected with respect to other error sources as errors caused by possible atmospheric parameter uncertainties similar to those reported in Table~\ref{Tab_error2}.
        
           \begin{table}[!t]
        \caption{Typical $internal$ uncertainties (sulfur abundance scatters, in dex) for different stellar types and S/N bins.
        These numbers refer to a single line in one spectrum. Since the final reported abundances are obtained
        by averaging up to three S{\sc i} lines and several $repeat$ spectra (when available), the actual relative uncertainties
        from star to star are much smaller.  Note: Only the S{\sc i}~675.7~nm line can be safely analysed in cool giants.}
        \label{Tab_error1}
        \centering
        \begin{tabular}{l r r r r r r}
            \hline
            & \multicolumn{2}{c}{Cool giant} & \multicolumn{2}{c}{Cool dwarf} &  \multicolumn{2}{c}{Hot dwarf} \\
            S/N   & $\sim$40 & $\sim$200  & $\sim$40 & $\sim$100  & $\sim$80 & $\sim$140\\
            \hline  
         S{\sc i}~674.3 &  --  &  --  & 0.06 & 0.06 & 0.05 & 0.02 \\
         S{\sc i}~674.8 &  --  &  --  & 0.04 & 0.03 & 0.02 & 0.02 \\
         S{\sc i}~675.7 & 0.03 & 0.03 & 0.06 & 0.04 & 0.02 & 0.02 \\
            \hline
        \end{tabular}
    \end{table}

        \section{AMBRE catalogue of sulfur abundances}
        \label{AMBRE_sulfur_catalogue}

\begin{table*}[!t]
        \caption{AMBRE catalogue of LTE sulfur abundances.}
        \label{TabAbund}
        \centering
        \begin{tabular}{c c c c c c c c c c c c }
            \hline
            \hline
                Star ID\tablefootmark{a}  & S/N & \T~(K) & \g & \meta & \alfaFe & $N_{6743}$ & $N_{6748}$ & $N_{6757}$ & \SH & $\sigma_{\SH}$ \\
            \hline
105332908999068032 &68 &5711&4.42& 0.30&-0.02&0& 2& 1& 0.31&0.01 \\
1153682508388170112&26 &6100&4.11&-0.08& 0.01&9&16&20& 0.05&0.04 \\
1155587858959873024&290&5795&4.04&-0.83& 0.27&0& 2& 1&-0.55&0.07 \\
1160260989536170880&220&5743&4.35& 0.02&-0.05&0& 1& 4& 0.07&0.01 \\
1160956465000504448&62 &5426&4.55& 0.04& 0.03&0& 1& 2&-0.01&0.11 \\
1167602394315412864&94 &5619&3.69&-0.95& 0.33&6&12&17&-0.56&0.03 \\
1174143182830505984&69 &5744&4.68& 0.17& 0.06&0& 1& 1& 0.10&0.07 \\
 ... & ... & ...  & ... & ... & ... & ... & ... & ... & ... & ... \\
 ... & ... & ...  & ... & ... & ... & ... & ... & ... & ... & ... \\
            \hline
        \end{tabular}
        \tablefoot{
        The full electronic table is available at the CDS.\\
        \tablefoottext{a}{$Gaia$ DR2 ID except HD-name for six very bright stars not present in the $Gaia$ second data release.}
        }
\end{table*}

\begin{table*}[!t]
        \caption{AMBRE LTE sulfur abundances  of the $Gaia$ benchmarks stars adopting their recommended atmospheric parameters.}
        \label{AbundBench}
        \centering
        \begin{tabular}{c c c c c c c c c c c c c }
            \hline
            \hline
                Star & Gaia DR2 or HD & S/N & \T~(K) & \g & \meta & \alfaFe & $N_{6743}$ & $N_{6748}$ & $N_{6757}$ & \SH & $\sigma_{\SH}$ \\
            \hline
 & & & & & & & & & & & \\
{\it Solar-type stars} & & & & & & & & & & & \\
$\alpha$ \ Cen B  & HD128621            & 45     & 5231 & 4.53 & 0.22 & 0.04 & 39   & 239  & 1013 & 0.25 & 0.25 \\
$\tau$ \ Cet      & 2452378776434276992 & 89     & 5414 & 4.49 &-0.49 & 0.23 & 615  & 1406 & 2505 & -0.38& 0.03 \\
Sun/Vesta         &   ---               & >1000  & 5771 & 4.44 & 0.00 & 0.00 & 1    & 1    & 1    & 0.00 & ---  \\
$\alpha$ \ Cen A  & HD128620            & 80     & 5792 & 4.31 & 0.26 & -0.02& 11   & 12   & 17   & 0.17 & 0.01 \\
18 Sco            & 4345775217221821312 & 79     & 5810 & 4.44 & 0.03 & 0.02 & 1854 & 1863 & 2281 & 0.00& 0.04 \\
HD 22879          & 3250489115708824064 & 124    & 5868 & 4.27 &-0.86 & 0.33 & 12   & 17   & 20   & -0.57& 0.01 \\
$\mu$ \ Ara       & 5945941905576552064 & 62     & 5902 & 4.30 & 0.35 & 0.00 & 626  & 1482 & 2849 & 0.14 & 0.02 \\
 & & & & & & & & & & & \\
{\it FGK subgiants} & & & & & & & & & & & \\
$\delta$ \ Eri    & 5164120762332790528 & 76     & 4954 & 3.76 & 0.06 & 0.04 & 0    & 18   & 275  & 0.12 & 0.07 \\
$\beta$ \ Hyi     & 4683897617108299136 & 102    & 5873 & 3.98 &-0.04 & -0.02& 2601 & 1162 & 2762 & -0.17& 0.01 \\
$\beta$ \ Vir     & 3796442680947600768 & 89     & 6083 & 4.10 & 0.24 & -0.13& 236  & 204  & 249  & 0.12 & 0.01 \\
 & & & & & & & & & & & \\
{\it F dwarfs} & & & & & & & & & & & \\
Procyon           & HD61421             & 128    & 6554 & 4.00 & 0.01 & -0.04& 2662 & 4844 & 5681 & 0.00 & 0.01 \\
HD 49933          & 3113219383954556416 & 105    & 6635 & 4.20 &-0.41 & 0.04 & 300  & 164  & 470  & -0.49& 0.03 \\
 & & & & & & & & & & & \\
{\it Cool giant} & & & & & & & & & & & \\
$\epsilon$ \ Vir  & 3736865265439463424 & 64     & 4983 & 2.77 & 0.15 & -0.07& 0    & 2    & 5    & 0.12 & 0.04 \\
 & & & & & & & & & & & \\
         \hline
        \end{tabular}
        \tablefoot{The electronic table is available at the CDS.}
\end{table*}

The final AMBRE catalogue of mean sulfur abundances is reported in Table~\ref{TabAbund}; a full version
is available in electronic form. 
Having applied all the rejection criteria of Sect.~\ref{outliers} ($repeat$ spectra with departing stellar
parameters, high-rotating stars, too low S/N spectra and non-detected lines) 
and the above averaging procedures, we finally 
provide the sulfur abundances with the associated dispersion for 1,855 different stars.
Among them, about 10\% are giant stars and the studied metallicity domain varies from $\sim$-1.9~dex
to $\sim$+0.7~dex.
These numbers have to be compared to the 719 stars with sulfur abundance of CS20 and to the 1,301 sulfur abundances of \cite{GES17}. The AMBRE-sulfur catalogue is therefore the largest ever published.
Moreover, it covers a metallicity range larger than any other,  particularly in the metal-poor regime
since almost 30 stars have \meta < -1.0~dex.

\begin{figure}
\resizebox{\hsize}{!}{\includegraphics{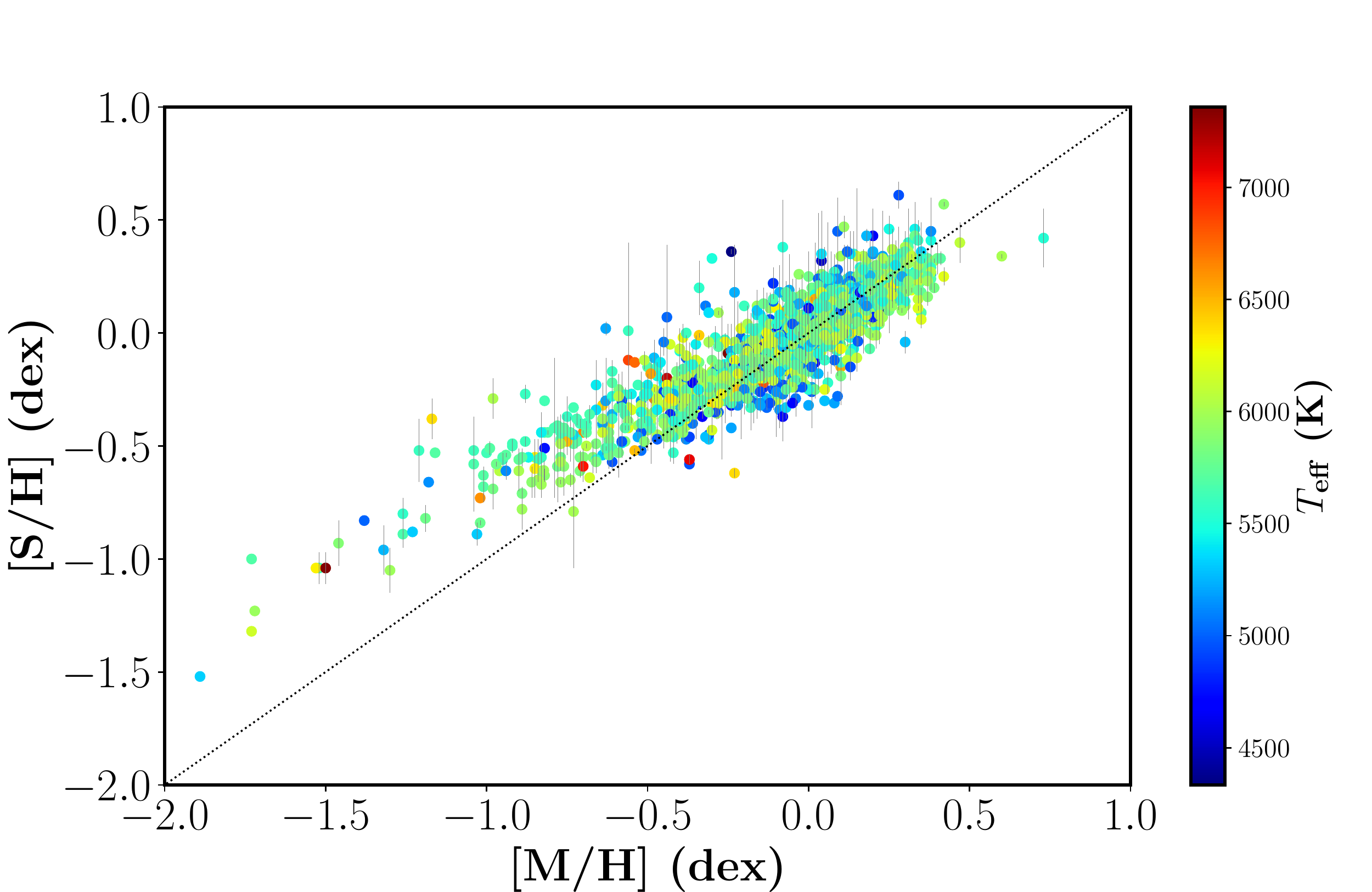}}
\caption{AMBRE-sulfur abundances \SH \ as a function of the mean stellar metallicity \meta \ for the whole sample. 
Error bars are the line-to-line dispersions listed in Table~\ref{TabAbund}. The stars without error bars
        are those with only one measured sulfur line (the 675.7~nm transition), hence no line-to-line scatter can be estimated.}
\label{Fig.S_H}
\end{figure}

In Table~\ref{TabAbund} the stars are identified by their $Gaia$ DR2 ID, except for six that only have a HD name.
This table contains the mean atmospheric parameters (\T , \g , \meta, \alfaFe) together with the mean 
S/N\footnote{Estimated with the AMBRE pipeline during the parametrisation process, i.e. at a spectral resolution $R \sim 15,000$.} of the spectra kept
when computing the mean final sulfur abundances. We recall  that these parameters were averaged by adopting exactly the same $WM$ procedure as for the abundances (see previous section). 
For a given star we also provide the number of analysed sulfur lines for each component of   multiplet~8 
($N_{6743},N_{6748},N_{6757}$) that have  finally been considered when computing its sulfur abundance.
These numbers range from unity (only one spectrum available) to several thousands when a large number of
${repeat}$ spectra were kept. The largest numbers of individual abundances derived for a 
given star are   5\,194, 4\,616, and 5\,636 for the three sulfur components, respectively.
We also report 1\,165, 1\,426, and 1\,855 stars with at least one measurement of the 674.3, 674.8, and 675.7~nm line,
respectively. Moreover, 1,049 stars have the three sulfur lines measured in at least one of their spectra.
Finally, as an indicator of the measurement uncertainty, we also report the dispersion ($\sigma_{\SH})$ between 
the different measurements   obtained for a given star (equivalent to a line-to-line 
scatter between lines and/or repeats). We note that this scatter can differ from the individual line error
$\Delta \SH$ described in the previous section.
The mean of these scatters is equal to 0.06~dex, and a dispersion lower than 0.02 and 0.05~dex
is found for 29\% and 57\% of the sample stars, respectively. We can 
therefore safely conclude that the reported sulfur abundances are very precise and consistent 
with each other. The derived sulfur abundances are 
shown in Fig.~\ref{Fig.S_H} as a function of the mean stellar metallicity. For positive metallicities it can be seen that
the sulfur abundances behave, in a first approximation, almost as \meta. However, \SH \ becomes increasingly higher than \meta \ for negative metallicities
down to \meta $\sim$ -1.0~dex, after which \SH-\meta \ stays almost
constant down to \meta $\sim$ -2.0~dex. This behaviour
is very similar to that of the \alfa-elements, suggesting that sulfur belongs to this class of chemical
species (see discussion in Sect. \ref{S=alpha}).

\begin{figure}
\resizebox{\hsize}{!}{\includegraphics{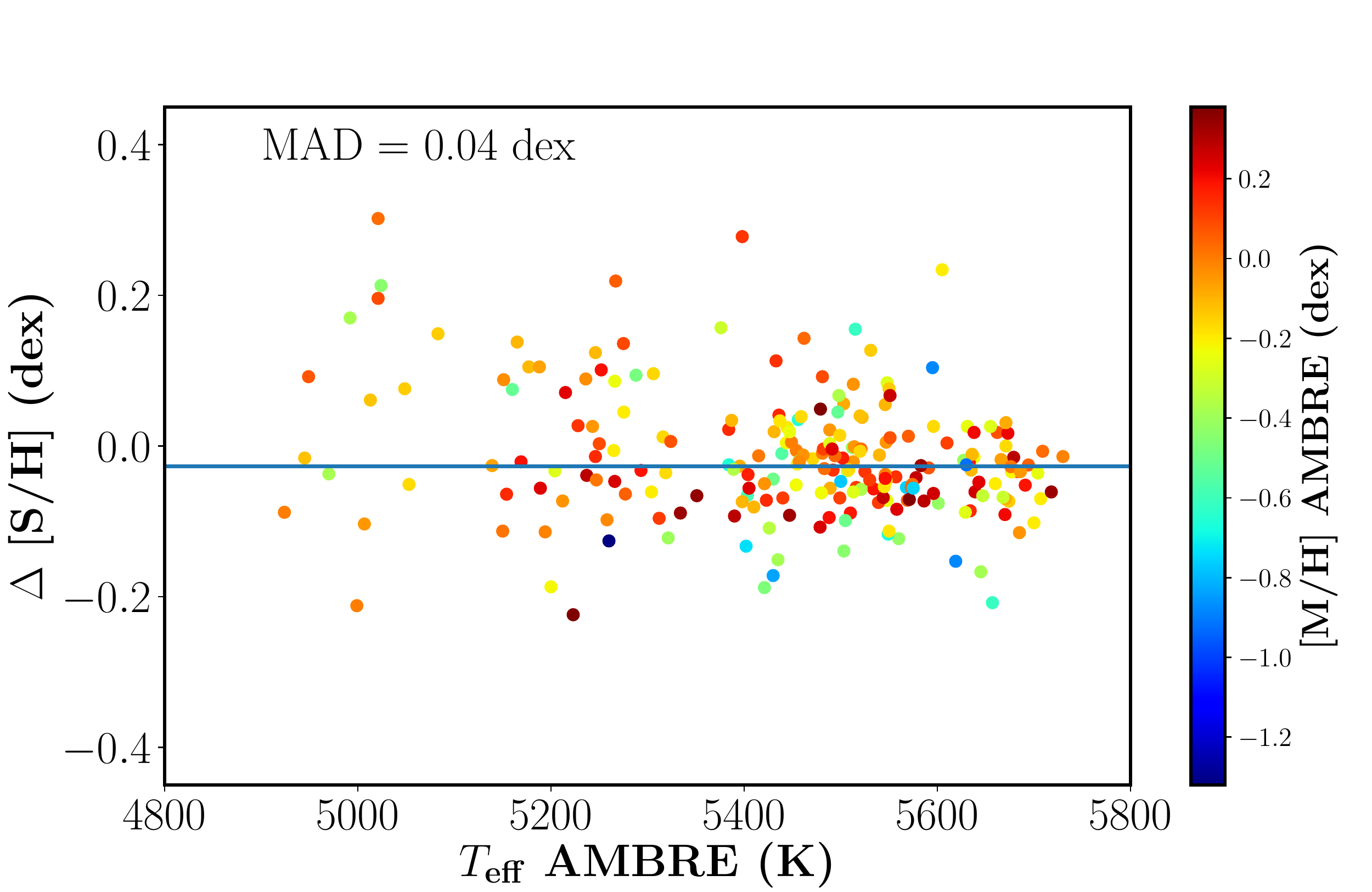}}
\caption{Comparison between the AMBRE and \cite{Porto20} sulfur abundances. The ordinate axis refers to the difference (\SH$_{\rm AMBRE}$ - \SH$_{\rm literature}$). The blue horizontal line
indicates the median of these differences (-0.027~dex), and the associated median absolute deviation is shown in the upper left corner. This MAD is equal to 0.03~dex if we only consider stars warmer than 5400~K (see text).}
\label{FigPorto}
\end{figure}

As a quality check of our derived sulfur abundances, we first verified that no systematic trends
are present between the AMBRE \SH \ and the effective temperature (over the range 4,500 -- 6,500~K) or the surface
gravity (which varies between 1.5 and 5.0~\gunits). We also compared our complete sample
of sulfur abundances (i.e. without selecting the best ones as done below) with those of CS20 estimated from the HARPS spectra of solar-type stars (see Fig.~\ref{FigPorto}). This catalogue contains a very large number  of stars (223) in common with ours. This large
number is explained by the fact that both samples
contain several HARPS spectra collected over similar epochs (most of them  probably identical spectra). The median of the differences
between the literature and the AMBRE values of  \SH \ is insignificant (-0.027~dex) with a very small associated dispersion, the MAD and the standard deviation being equal to 0.04~dex and 0.08~dex respectively (see Fig.~\ref{FigPorto}). We note that a large part of these small differences can be explained by the different atmospheric parameters adopted in the two studies.
For example, the dispersion associated with the differences in \T, \g,  and \meta\ are 63~K, 0.14, and 0.05~dex, respectively. It can  be seen, however,  that
the differences in \SH \ seem to slightly increase
towards cooler stars, revealing perhaps some differences in the analysis (possibly
caused by different considerations of some molecular blends?). 
%For example as far as we know, no molecular lines were included in the analysis of CS20 and this could affect their derived \SH \ for the coolest stars. 
Moreover, CS20 seem to favour in their discussion their stars with effective temperatures within
$\pm$500~K around the solar value, the range in which their errors are smaller. For the stars in common and
having \T $\ge$ 5400~K, the agreement is indeed better since 
the MAD and the standard deviation decrease to 0.03~dex and 0.06~dex, respectively (the median
of the differences staying insignificant).
As a consequence, we can safely conclude that the agreement between the two independent analyses is very satisfactory, and this
confirms the high accuracy of the AMBRE catalogue of automatically derived sulfur abundances. 

Finally, we provide in Table~\ref{AbundBench} new sulfur abundance of the $Gaia$ benchmark stars
present in our sample (several of them having a very large number of available spectra). 
These FGK-type stars have carefully studied atmospheric parameters
derived using different analysis techniques, and they are thus very helpful to calibrate and/or validate large spectroscopic
surveys. Although these stars are already present in Table~\ref{TabAbund}, their AMBRE
atmospheric parameters may differ slightly   from the commonly accepted values. We  therefore
recomputed their sulfur abundances from all their available spectra, assuming for each of them the accepted parameters summarised in \cite{Benchmark18} (where the adopted \alfaFe \ is the mean of the individual Mg, Si, Ca, and
Ti abundances). We note that, for the Sun, we analysed a very high-S/N HARPS spectrum of Vesta. 
These new \SH \ values in Table~\ref{AbundBench} differ only marginally (by a few hundredths of dex)
from those of Table~\ref{TabAbund} and this only results from the small differences in the adopted
atmospheric parameters. These new abundances should   be favoured in any future calibration or validation
studies of sulfur abundances.

%  Tests calibation
% 1 - Spectre Soleil /Vesta (5772/4.44/0.0/0.0)
%--> Vesta: Raie1/2/3=-0.07/+0.04-0.08?/-0.03 => [S/Fe]~-0.04
%--> SunHinkle: Raie1/2/3=-0.1/+0.02-0.08?/-0.06 => [S/Fe]~-0.05
%
%--> Arcturus/AMBRE+Hinkle: spectres jetés. Pb normalisation??
%
%--> Procyon: pb largeur raies? Vmacro pas pris en compte bien que R=40,000
%      Raie1/2/3=-0.06/-0.01-0.08?/-0.03 => [S/Fe]~-0.05
        
        \section{Behaviour and evolution of sulfur in the Milky Way}
\label{Discussion}

In this section we analyse the catalogue presented above by selecting stars with the best
measured sulfur abundances. 
For this  purpose we use the following:  
\begin{itemize}
\item For \meta $\ge$-1.0~dex: We selected stars having a sulfur abundance dispersion $\sigma_{\SH}$ lower than 0.05~dex
(i.e. those having at least two lines measured consistently).
\item In the same metallicity regime: We kept the other stars having only the 675.7~nm
component, but measured in a high-quality spectrum (S/N > 100). 
\item For the metal-poor regime (\meta<-1.0~dex):  Because fewer  stars are available, we slightly relaxed these 
strict criteria by adding any stars having spectra with S/N  higher than 50. 
\end{itemize}
The resulting subsample
consists of 1,203 stars with very high-quality sulfur abundances (65\% of the whole AMBRE-sulfur catalogue, see top panel of Fig.\ref{Fig.S_H-Best}).\\

We also built a {\it golden} sample by selecting stars with even better sulfur abundances and satisfying the following criteria: 
\begin{itemize}
\item 
They should have at least three measurements of each individual S{\sc i} line (i.e. with at least nine
 abundances available to estimate their mean sulfur abundance) and a total dispersion $\sigma_{\SH}$ lower than 0.05~dex for their mean [S/H]. 
\item We also included
metal-poor stars (\meta<-1.0~dex) having fewer measured lines, but spectra with a S/N higher than 150. 
\end{itemize}
This $golden$ sample contains 
540 stars with extremely high-quality and precise sulfur abundances. They are shown
in the three bottom panels of Fig.\ref{Fig.S_H-Best}. The properties of
these best stars are described in more detail below, but the same
conclusions are reached with the largest sample of  1,203 stars, although the slightly larger dispersion could slightly blur  some of the figures
shown in the following.
 Finally,  we   discuss the sulfur abundances with respect to the mean metallicity
([S/M]) obtained by substracting the [S/H] estimated in the present work from the mean metallicity
\meta \ previously derived within the AMBRE Project ([Fe/H] abundances estimated from individual iron
lines being not available for the whole sample).

\subsection{Sulfur as an \alfa-element}
\label{S=alpha}

\begin{figure}
\resizebox{\hsize}{16cm}{\includegraphics{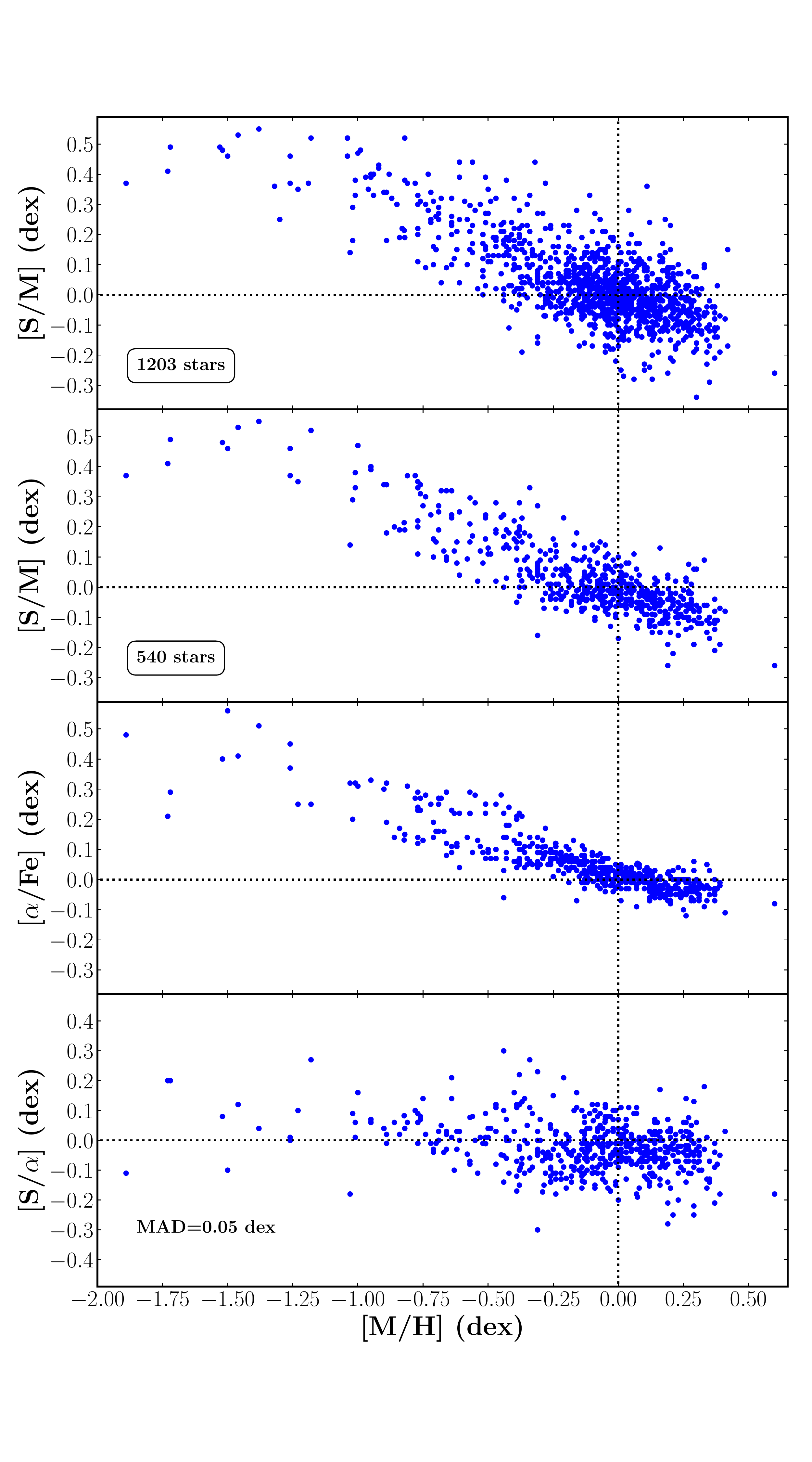}}
        \caption{Ratio of sulfur abundance to mean metallicity   \SM \ 
        as a function of the mean stellar metallicity \meta \ for the best derived abundances. Top two panels: Dispersion smaller than 0.05~dex and/or high-S/N spectra (top panel) and the same criteria plus a selection of stars having at least three measurements of the three
individual S{\sc i} lines ({\it golden} sample, second panel from top). Bottom two panels:  Behaviour of \alfaFe \ and [S/\alfa]\ vs. mean metallicity
for the {\it golden} sample.}
\label{Fig.S_H-Best}
\end{figure}

\begin{figure}
\resizebox{\hsize}{!}{\includegraphics{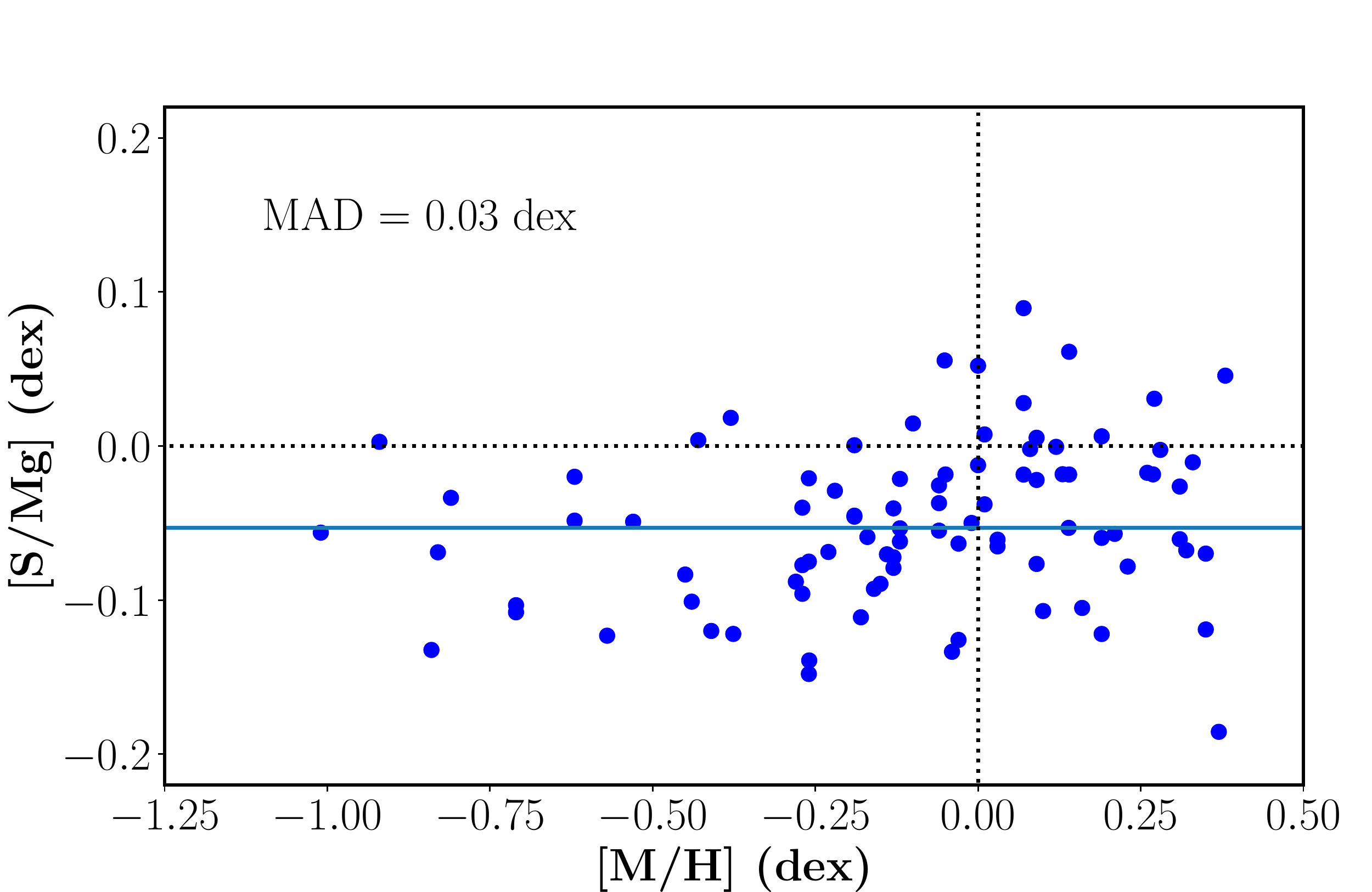}}
        \caption{Ratio of sulfur to magnesium abundance [S/Mg]  
        as a function of the mean stellar metallicity \meta \ for stars in common with \cite{Pablo20}.
        The blue horizontal line indicates the median of  [S/Mg]  (-0.05~dex) over the whole metallicity domain and the associated median absolute deviation is reported in the upper left corner.}
\label{Fig.S-Mg}
\end{figure}

\begin{figure}
\resizebox{\hsize}{10cm}{\includegraphics{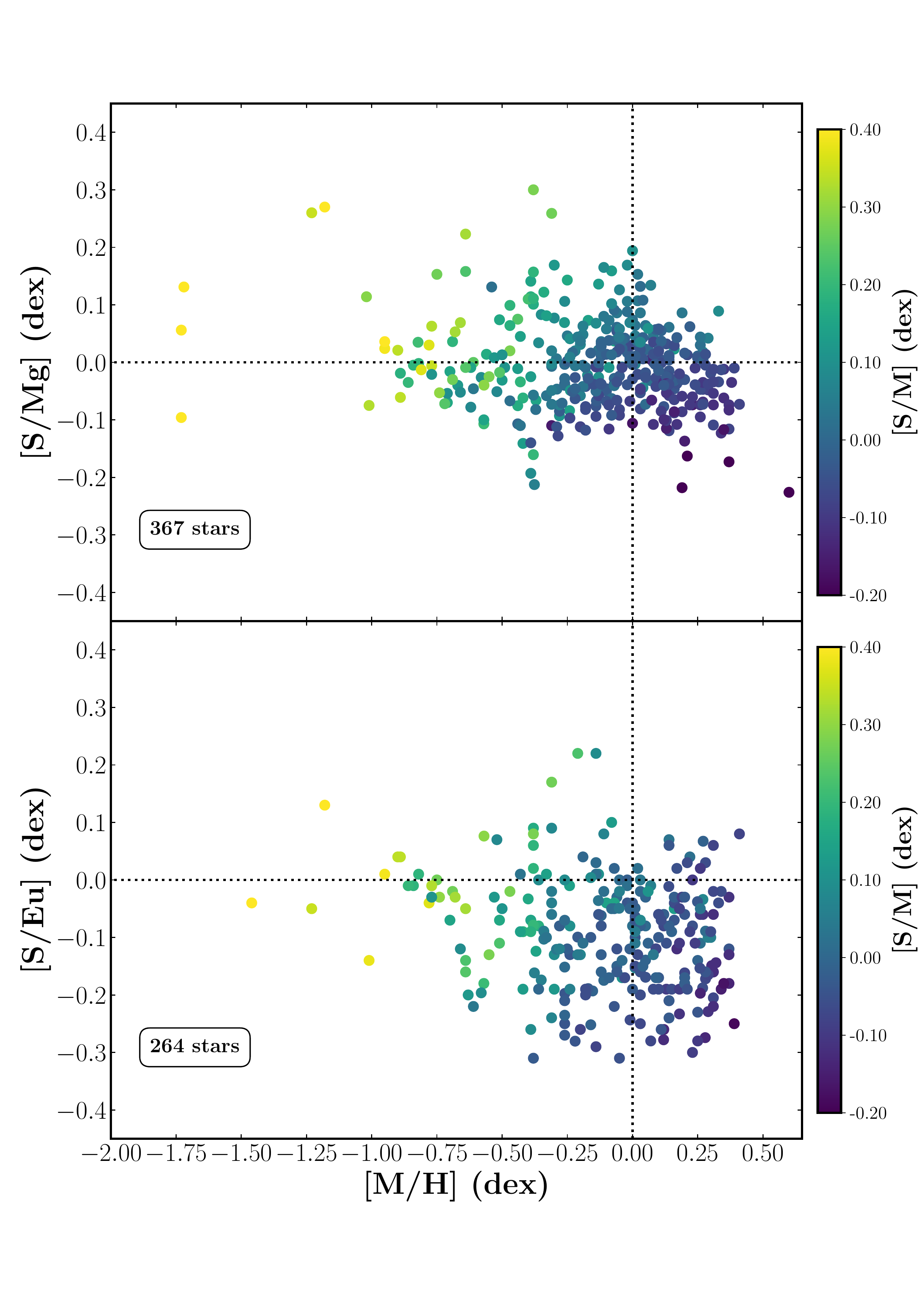}}
	\caption{Abundance ratios of sulfur to magnesium  ([S/Mg], top panel) and to europium ([S/Eu], bottom
        panel)  
        as a function of the mean stellar metallicity \meta \ colour-coded with [S/M],
        adopting Mg and Eu abundances derived by \citet[][top panel]{Sarunas17}
        and \citet[][bottom panel]{Guiglion18}.}
\label{Fig.S-MgEu}
\end{figure}

Because of its nucleosynthesis channel, sulfur belongs to the family of the \alfa-elements, as do  oxygen and
magnesium, for example. For these chemical species Milky Way evolution models predict a 
specific variation of the abundance ratio
 of \alfa-elements  to   iron  (\alfaFe) 
with respect to metallicity \meta.  
This is observationally confirmed for most \alfa-species, including
sulfur to a lesser extent (see the Introduction for   references). However, the present AMBRE-sulfur catalogue
offers the possibility to draw a more global and homogeneous picture of the sulfur abundance variations in the Milky Way
than the previous studies, thanks to its large statistics, accurate measurements, and large metallicity range covered.

As already shown in Fig. \ref{Fig.S_H} for the complete sample, and more clearly shown in 
Fig. \ref{Fig.S_H-Best} for the high-precision and {\it golden} samples exhibiting our very best sulfur abundances, 
the abundance ratio of sulfur to mean metallicity  ([S/M]) exhibits a
variation with respect to the mean metallicity that is very similar to that of the
mean \alfaFe \ ratio provided by the AMBRE Project
(third panel of Fig. \ref{Fig.S_H-Best}). The dispersion is smaller
for \alfaFe \ because it is derived from a much larger number of lines belonging
to different chemical species than the maximum of  three  sulfur lines studied in this work. This similar behaviour between sulfur and \alfa-elements 
is also clearly illustrated
in the bottom panel of Fig. \ref{Fig.S_H-Best} where the mean [S/\alfa] value
is equal to zero with a very small dispersion over the whole metallicity range (MAD=0.05~dex).

On the other hand, the suspected plateau (see Introduction) of the metal-poor regime (Galactic halo stars with \meta $\la$-1.0~dex) 
can be studied thanks to the 27 stars analysed in this work (only those of the $golden$ sample
are shown in Fig. \ref{Fig.S_H-Best}). This number of stars is much larger than the numbers reported by other large surveys of sulfur abundances (only two stars in CS20 and about
half a dozen in \citealt{GES17}). It should be noted that 
%this  plateau is found at
the mean [S/M] in the metal-poor regime is found at +0.44~dex, a classical level for \alfa-elements. The variation of [S/M] with metallicity for \meta $\la$-1.0~dex
is compatible with a flat
behaviour, as confirmed by a locally weighted scatterplot smoothing (LOWESS) fit of the data \citep[][]{LOWESS}. 
However, a rather large dispersion (close to 0.15~dex) around this mean can be seen, probably caused by
the difficulty in  measuring the sulfur lines in this low-metallicity domain in spectra with S/N that can be as low as 50 for the high-precision sample.
Nevertheless, our study therefore invalidates the suggestion of a steady increase 
in [S/M] with decreasing metallicities (see   discussion in the Introduction).
However, the ratio [S/\alfa] seems to become slightly positive (although
with a large dispersion; see Fig. \ref{Fig.S_H-Best} bottom panel) for these low metallicities: 
the mean [S/\alfa] for \meta $\la$-1.0~dex is close to 0.07~dex with a dispersion MAD=0.065~dex. 
This dispersion could result from the analysis, but 
%caused by the difficulty to measure the sulfur lines in this low-metallicity domain
%in spectra with SNR that can be as low as 50.
it could also reflect the fact that the lines of different chemical species adopted to derive the \alfa-abundances 
could differ from one metallicity regime to another. As a consequence \alfaFe \ could more or less correlate with [Mg/Fe], for example, depending
on \meta. We note, however,   that the dispersion in \alfaFe \ looks larger than in [S/M]
in this low-metallicity regime. Such a dispersion could therefore reveal
the heterogeneity nature of the Galactic halo.

We now focus on the supersolar metallicity regime (\meta$\ga$0) where the values of  [S/M] and \alfaFe \  shown
in Fig. \ref{Fig.S_H-Best} become negative. This  decrease is  extensively discussed in
\citet{Pablo20} for magnesium and, independently, clearly seen  again here for sulfur.
This continuous decrease is not seen in the sulfur sample of
CS20, and not always seen in other Galactic studies of \alfaFe \ behaviours.
% estimated by large spectroscopic surveys (as APOGEE, GES, ...)
%but we do believe that it is real. First, 
We recall that \citet{Pablo20}  show  that the flattening reported 
by some previous studies is an artefact created by an incorrect continuum normalisation procedure in crowded-line spectra 
of very metal-rich stars.
We also note that this decrease for sulfur and magnesium is in perfect agreement
with Galactic evolution models. There is no reason why the production rate of \alfa-elements
(and sulfur) would suddenly increase around solar metallicity and/or why the iron production 
would be constant or smaller to produce an almost constant \alfaFe \ ratio at high metallicity. 
For instance, \citet{Palla20} predict an \alfaFe$\sim$-0.2~dex at \meta$\sim$+0.5~dex
in close agreement with our observations \citep[see also the models of][]{Nikos18, Romano10, Koba20}.
% Such non-optimized procedures would lead to overestimated \alfaFe \ ratios above \meta$\sim$0,
%in strong disagreement with expected behaviours. On the contrary, the smaller AMBRE \alfaFe \ and [S/M] ratios
%can be more easily interpreted by Galactic evolution models.
% Palla2020: However, a word of caution is necessary when speaking ofthe APOGEE data relative to [Mg/Fe] at high [Fe/H] val-ues, since the [Mg/Fe] ratio could have been overestimated,as suggested by J ̈onsson et al. (2020, submitted)

Then, we   compared the AMBRE-sulfur abundances with magnesium
abundances previously derived within the AMBRE Project.
We first show in Fig. \ref{Fig.S-Mg} stars having accurate Mg abundances derived in \cite{Pablo20} from carefully selected lines
and for which the normalisation procedure
was optimised for the high-metallicity regime. To increase the size of the comparison sample, we slightly relaxed the quality criterion of 
the selected sulfur abundances (dispersion lower than 0.1~dex) and this resulted in
89 stars in common between both catalogues. Again, the two \alfa-elements Mg and S show the same overall behaviour
with metallicity.
%Stats pour S/Mg_Pablo pour ces étoiles
%mediane, MAD= -0.053139741454545454 0.03471647545454544
%Pour [M/H]<=-0.25: mediane, MAD= -0.0761788762254902 0.03934383772549024
%Pour [M/H]>=-0.10: mediane, MAD= -0.02554060400000001 0.034086643999999985
%Stats pour alpha/Mg_Pablo pour ces étoiles
%mediane, MAD= -0.0033378146999999995 0.01762010857272728
%Pour [M/H]<=-0.25: mediane, MAD= -0.0527694 0.024978257000000004
%Pour [M/H]>=-0.10: mediane, MAD= 0.0033944649999999993 0.008551866
%{\bf METTRE LE MAD de Mg/alpha como es alpha/Mg para
%las metal-poor? Está centrado en cero o es negativo?
%Porque si el [S/alpha] es positivo para las metal-poor,
%probablemente es el [Mg/alpha] que está negativo.}. 
The median [S/Mg] ratio over the whole metallicity range is close to -0.05~dex with 
an extremely small dispersion (MAD=0.03~dex). This median is close to -0.08~dex and -0.03~dex for \meta<-0.25~dex
and \meta>-0.1~dex (with similar small dispersions), respectively. We also note that,
for these metallicity regimes, the median of [\alfa/Mg] is equal to
-0.05~dex and 0.0~dex, respectively, leading to  
the slightly negative mean [S/\alfa] ratio shown in Fig.~\ref{Fig.S_H-Best} when \meta $\gg$~-1.0~dex. 
Such [S/Mg] ratios
could be caused by different calibrations adopted for the abundance
derivations (although the procedures adopted by these studies are very similar), but could also be real since the production rates (yields) of these two elements could
be slightly different. Our slightly negative [S/Mg] ratio in the metal-poor regime seems to differ from
the positive [S/Mg] value predicted
by \citet[Fig.5]{Koba20} and \citet[Fig.13]{Nikos18}, whereas the agreement
seems much better for metal-rich stars. 
Finally, the variation of [S/Mg] with metallicity (higher at higher \meta) could suggest that these two elements do not 
vary in perfect lockstep during the whole Galactic evolution (but see Fig.~\ref{Fig.S-Ages}, middle panel).

We  compare in the top panel of Fig.\ref{Fig.S-MgEu} the sulfur-to-magnesium abundance
ratio with respect to the metallicity adopting the magnesium abundances
from the study of \cite{Sarunas17}.
We considered the best derived magnesium abundances of this work 
by rejecting those with too large errors, and found 367 stars in common.
We also show in the bottom panel of Fig. \ref{Fig.S-MgEu} a similar plot
showing the sulfur-to-europium abundance ratio thanks to data derived by 
\cite{Guiglion18}. Here, we selected the best Eu abundances derived
from at least two lines and having a small dispersion (264 stars are found
in common with the AMBRE-sulfur sample). From these two plots it can be
seen that sulfur, magnesium, and europium are closely correlated and follow a  similar
behaviour: the abundance ratios seem to be rather constant with the metallicity. 
More precisely, the top panel of Fig. \ref{Fig.S-MgEu} is very similar to Fig. \ref{Fig.S-Mg},
although the dispersion is larger, probably resulting from the different
methodoloy adopted by \cite{Sarunas17} for deriving Mg abundances.
Regarding the [S/Eu] ratio, it also seems to stay constant with metallicity,
although sulfur could be slightly underabundant (by about $\sim$0.15~dex) with respect to europium at any metallicity.
%larger Mg and Eu abundances
%are always encountered at larger [S/M]. This is even clearer in the metallicity
%range -1.0$\la$[M/H]$\la$-0.5~dex, where the thin/thick disc chemical dichotomy is usually found
%and, where [S/M], [Mg/Fe] and [Eu/Fe] are nicely correlated.
Thus, since magnesium and europium are believed to be predominantly
produced in Type II supernovae (although other production sites
are invoked for Eu), Fig. \ref{Fig.S-MgEu} again confirms
that sulfur could be predominantly produced by the same nucelosynthesis
channel in SN~II.

In summary, and as a conclusion of this subsection, we can therefore safely state that the $\alpha$  nature of sulfur is clearly confirmed by the AMBRE
Project over a very large metallicity domain.

\subsection{Kinematic and orbital properties}
\label{Sect.kinematics}
\begin{figure}
\resizebox{\hsize}{18cm}{\includegraphics{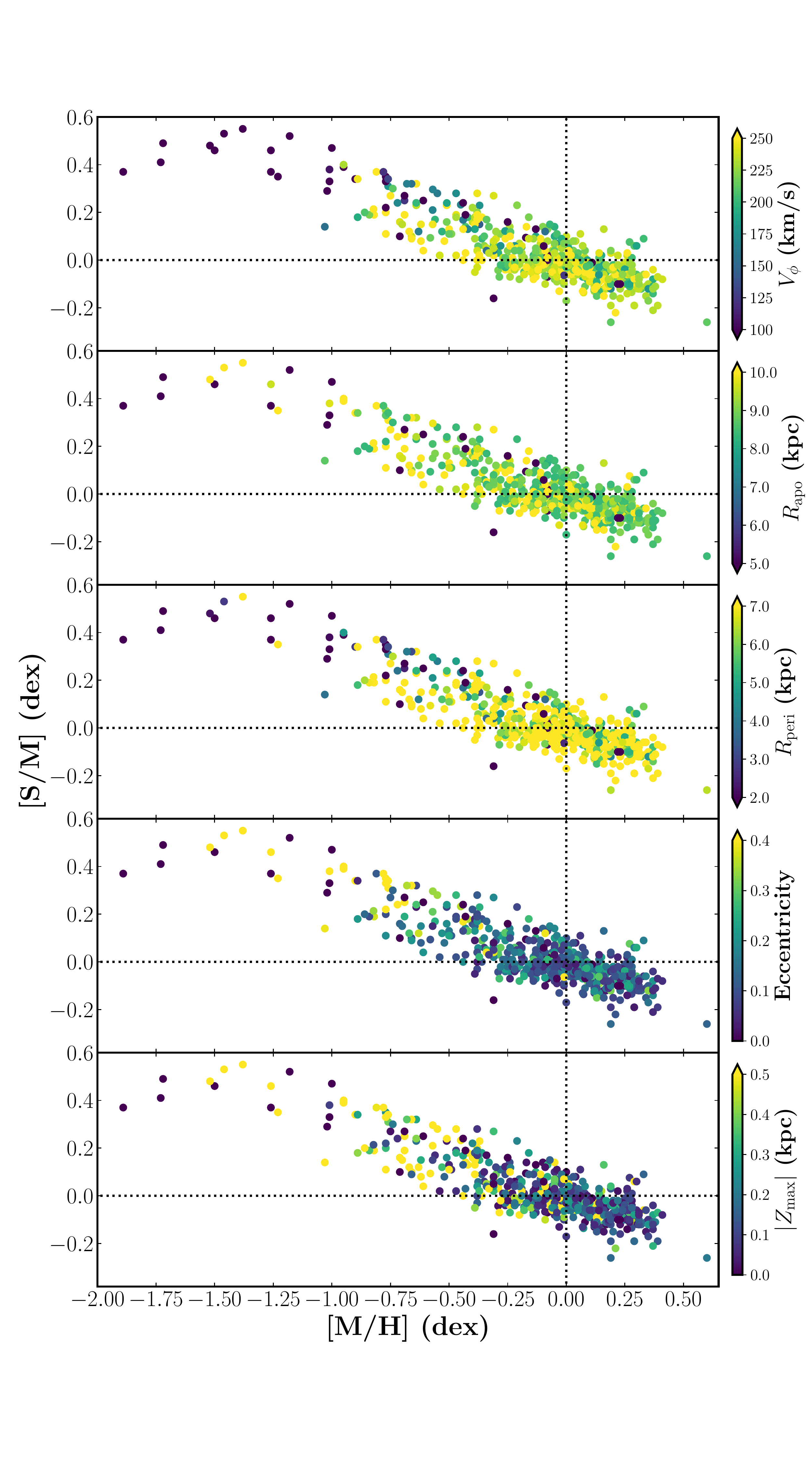}}       \caption{Ratio of sulfur to mean metallicity \SM \ as a function of the mean metallicity \meta \ colour-coded (from top to bottom) by the rotational velocity
        ($V_{\rm \phi}$), the  
        pericentre and apocentre radii
        ($R_{\rm peri}$ and $R_{\rm apo}$),
	the eccentricity, and the maximum height above the Galactic plane ($Z_{\rm max}$). }
\label{Fig.kinematics}
\end{figure}

Thanks to Gaia/DR2 astrometry \citep{gai18} and associated distances from \cite{Coryn18}, 
we   computed the Galactic cartesian coordinates of our sample stars. As expected, our sample is predominantly composed of stars located in the solar vicinity: about 90\% of them are found within 200~pc of the Sun. Moreover, except for a few cases that are found at a distance from the Galactic plane
up to $\sim$1~kpc, most stars are located within a distance smaller than 400~pc from the plane.
Their kinematic properties were then estimated thanks to the AMBRE radial velocities,
assuming that the Sun is located in the Galactic plane  at 8.2~kpc from the centre and 
adopting the velocities reported by \cite{Schonrich10} for its motion with respect to the local standard of rest. 
Finally, orbital parameters
were calculated with the $galpy$ code \cite{Bovy15}.
We refer to Santos-Peral et al. (2021)
for a detailed description of the methodology adopted for computing these orbits.

The complex chemo-dynamical characteristics of the disc stellar populations are still a matter of debate, and are constantly being updated thanks to progressively more complete samples inside and outside the solar neighbourhood. 
It is today largely admitted that the Galactic disc presents a chemo-dynamical bimodality, usualy described as the combination of two components: the thin and the thick disc. Since the discovery of this bimodality in stellar density distributions \citep{Yoshii82, Gilmore83} several studies have shown the kinematical \citep[e.g.][]{Bensby03, Reddy06, Georges17} and chemical \citep[e.g.][]{Vardan12, Recio14, Hayden15} distinctions between the two components. The thick-disc  presents a larger scale height and a shorter scale length, and it is kinematically hotter with respect to the thin-disc. It is also reported to be \alfaFe \ enhanced with respect to the thin disc at all metallicities. Finally, the stellar age distribution is older for the thick disc than for the thin disc. Different evolution models and simulations have tried to interpret the observed distributions. Although the Galactic disc evolution is still a matter of debate, it is generally accepted that the thick disc phase corresponds to the early disc component, and that about 8 Gyr ago a discontinuity in the disc evolution allowed the formation of the thin disc.
The study of the disc dichotomy using sulfur abundances has been hindered until now by the lack of good statistics and the low precision of the abundance estimates. In this section we take advantage of our  large sample of precise sulfur abundances in the solar neighbourhood to analyse the chemo-dynamical correlations in the [S/M] versus [M/H] plane.

We show in Fig.\ref{Fig.kinematics} how the sulfur abundances of the best sample 
defined earlier ({\it golden} sample, see  Fig.\ref{Fig.S_H-Best}) are 
related to some of their stellar orbital
properties such as the pericentre and apocentre radii
($R_{\rm peri}$ and $R_{\rm apo}$), 
the eccentricity ($e$), the rotational velocity ($V_{\rm \phi}$), and the maximum height above the  Galactic plane ($Z_{\rm max}$).

First, a gap can be suspected between the [S/M]-rich  and the [S/M]-poor populations, although its precise
location cannot be easily defined, as  is often the case for the other \alfa-elements. We note, however,  that 
the chemical separation of the two disc components seems to depend on the studied \alfa-element 
\citep[see e.g.][]{Sarunas17}.
Nevertheless, the kinematical and dynamical characterisation   confirms that the observed [S/M] dispersion at a given metallicity is not the result of the abundance uncertainties, clearly illustrating the above-described disc bimodality. This is particularly true for stars having metallicities lower than -0.5 dex. As is the case for other \alfa-species, the thick versus thin disc dichotomy is observed to be blurred at higher metallicities, and  is today a matter of debate \citep{Vardan12, Hayden17}.

More specifically, it can be clearly seen in  Fig. \ref{Fig.kinematics} that metal-poor sulfur-rich stars (MPSR, for metallicities found between [-1.0,-0.5]) have orbital properties
that strongly differ from the more metal-rich sulfur-poor ones (MRSP):
\begin{itemize}
\item First, regarding the rotational velocity, two disc components are present and
they differ in $V_{\rm \phi}$ and [S/M]. The less enriched sulfur stars (thin disc)
have lower rotational velocities close to the solar velocity. Only the most metal-poor stars have much 
lower rotational velocities. A gradient of $V_{\rm \phi}$ with \meta \ is also seen, as already mentioned for other \alfa-species \citep[see e.g.][]{Recio14, Georges17}.
\item Then, most of the MPSR are found in the inner Galactic regions contrarily to
the MRSP that are predominantly located close to the solar apocentrer and pericentrer radii.
                Radial gradients in both $R_{\rm peri}$ and $R_{\rm apo}$ can also be suspected 
                (for a given metallicity bin),
and they are correlated with the [S/M] enrichment. For instance, at
a given metallicity (for \meta $\ga$ -1.0~dex), stars having a smaller apocentre radius are 
the most enriched in sulfur
\citep[for a similar discussion, see][]{Hayden17}. 
If thick-disc stars are defined as being more sulfur-rich
(see Sect.~\ref{Sect.Popul}), it can be easily seen that the thick disc is
more radially concentrated than the thin disc.
\item The eccentricity of the MPSR is higher than the that  of the 
MRSP, and every star with a metallicity higher than $\sim$-1.0~dex is on a quasi-circular orbit ($e \la 0.15$). Only the most metal-poor stars with the highest [S/M] ratios are found on 
eccentric orbits,  as  is the case for halo stars, with a few exceptions 
(also detected in $Z_{\rm max}$) confirming the wide range of stellar orbits
in this metallicity range.
\item Finally, the bottom panel of Fig. \ref{Fig.kinematics} reveals
a settling of the disc stars for the MRSP, and the MPSR are mostly found
at several hundreds of parsecs above the Galactic plane, as already described in \citet{Hayden17}.
\end{itemize}
In summary, Fig. \ref{Fig.kinematics} shows that the two
disc components can be defined by studying both the kinematic
and the sulfur content. This confirms the high quality
of the sulfur abundances of the selected stars, and we   show 
in Sect.~\ref{Sect.Popul} that the thin and thick disc could also be
defined based solely on their sulfur content, as could be done by adopting
any other $\alpha$-elements.

Furthermore, Fig. \ref{Fig.kinematics} also reveals that all the metal-poor
(\meta $\la$ -1.0~dex) stars of the sample have much lower rotational velocities than the more metal-rich stars.
These $V_{\rm \phi}$ values lower than $\sim$150~km/s are typical of halo star members.
However, it can also be seen in the bottom panel of Fig. \ref{Fig.kinematics} that a large part
of them are located very close to the Galactic plane. This could reveal that these
halo stars are presently crossing the plane. Their large number could 
result from the complex selection function of the AMBRE catalogue revealing a possible bias
towards the identification of metal-poor stars in the solar neighbourhood.

\subsection{Sulfur abundances and stellar ages}

\begin{figure}
\resizebox{\hsize}{11cm}{\includegraphics{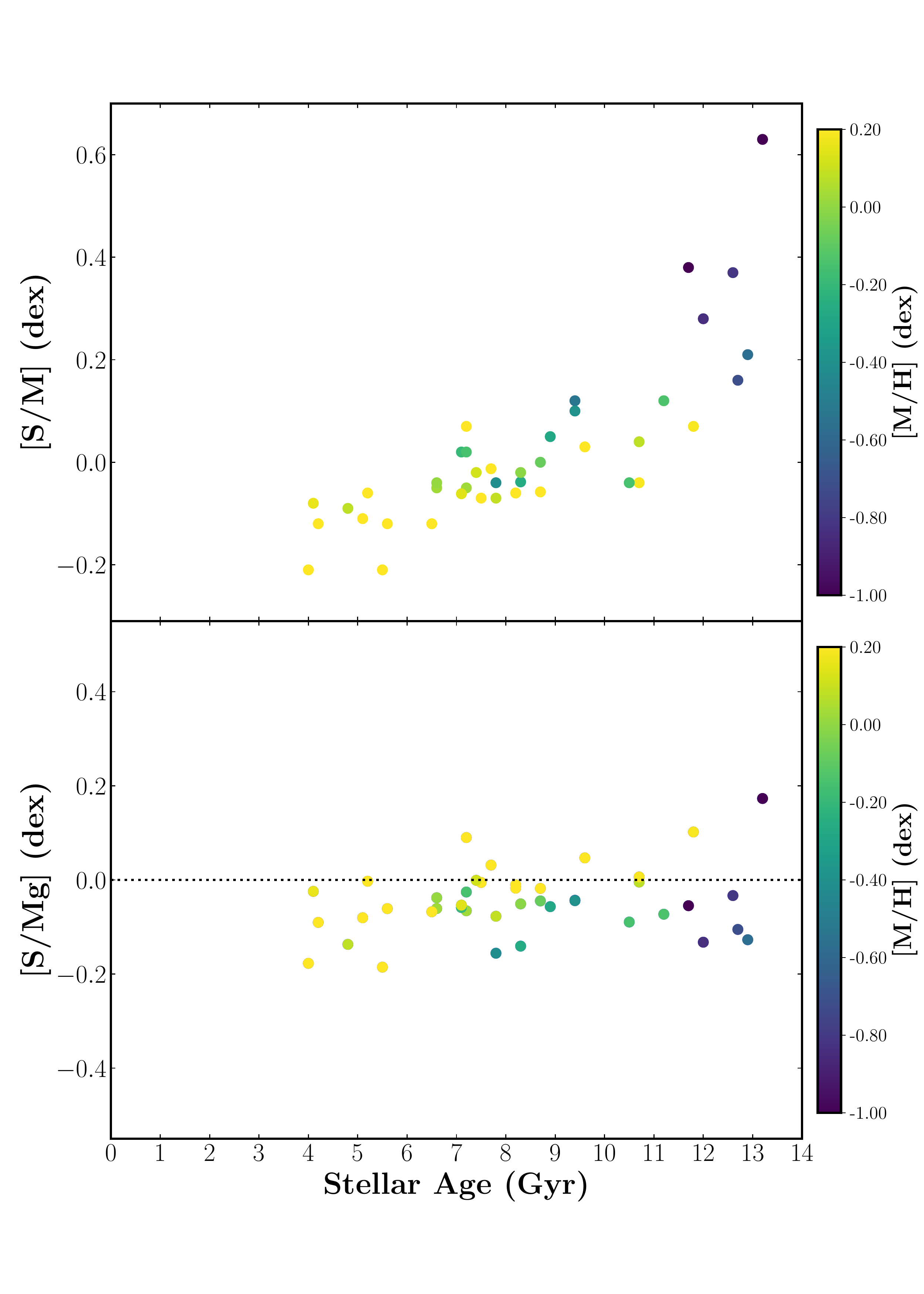}}
        \caption{Ratio of sulfur abundance to mean metallicity \SM \ (top panel) and to magnesium abundance [S/Mg] (bottom panel) as a function of the stellar ages for the main sequence turn-off and subgiant stars of the sample. The colour-coding corresponds to the
mean stellar metallicity.}
% \meta \ (bottom panel)  Symbols
%for the different Galactic populations are as in Fig.\ref{Fig.S-popul}.}
	\label{Fig.S-Ages}
\end{figure}

Santos-Peral et al. (2021, submitted) have estimated accurate and reliable ages for about 400 AMBRE stars
using an isochrone fitting method, as in \cite{Georges16}, and distances from \citet[][based on Gaia/DR2 parallaxes]{Coryn18}.
For this  purpose they selected only main sequence turn-off and subgiant stars for which age estimates are
more accurate since stellar ages increase quickly when stars cross these regions of the HR-diagram.
We refer to this article for a detailed description of the adopted methodology to derive the stellar ages adopted in the present study.

Among our {\it golden} sample of stars with the best derived sulfur abundances, about 10\% are
in common with Santos-Peral et al. (2021) and have ages with
rather small uncertainties (the mean of their age relative errors is equal
to 17\% with a dispersion of 9\%). The variation of [S/M] and [S/Mg] abundance ratios as a function of the stellar ages and colour-coded with the mean metallicity are
shown in Fig.~\ref{Fig.S-Ages}. The following  can be seen:
\begin{itemize}
\item The \SM \ ratio (top panel) of disc stars decreases towards younger 
and more metal-rich stars in a continuous way
from $\sim$10~Gyr to $\sim$4~Gyr and with a rather small dispersion. Moreover, this decrease is also
seen down to negative \SM \ values for ages younger than $\sim$5-6~Gyr. We clearly see the sulfur depletion with respect to iron
                in the solar vicinity for the youngest stars. We found a slope for the \SM \ versus age relation close to $\sim$0.02~dex/Gyr for stars younger than $\sim$10~Gyr (in very good agreement with the gradient reported by Santos-Peral, 2021, for magnesium).
This contradicts some previous claims where most thin-disc stars seem to have almost constant \SM \ ratios with age; for instance,
the slope in CS20 is two times smaller.
However, %we are confident in 
the behaviour shown in Fig.~\ref{Fig.S-Ages} is easy to interpret:
As expected by chemical evolution models,
the \alfaFe\  content of stars in the supersolar metallicity regime is believed to continue
to decrease with time (the production of \alfa-species by SNII is strongly reduced with respect to
that of iron-peak elements in SNIa). More explicitly, younger metal-rich
                stars should always have lower
\alfaFe \ (and hence \SM) ratios, as predicted by any chemical evolution model of the Galactic discs
\citep[for a recent reference, see Fig. 9 in][]{Palla20}.
\item The stars older than $\sim$8~Gyr 
        follow the same behaviour as younger stars: the older they are, 
the more sulfur-enriched and metal-poor they are. However, the increase in \SM \ with age is much faster and steeper. 
Moreover, all the oldest stars in our sample, which are the most metal-poor, are enriched in sulfur. Once again, this is in complete
agreement with chemical model predictions.
%\item the SRMR stars, on another hand, do not exhibit this trend: their high \SM \ ratio (inherent
%to their definition) surprinsignly stays almost constant for intermediate ages between $\sim$6~Gyr to $\sim$12~Gyr.
%{\bf ca s'interprete comment, ça?}
\item It can also be seen in the bottom panel of Fig.~\ref{Fig.S-Ages} that sulfur and magnesium
abundances stay close to each other whatever the stellar age (but see the above discussion
on this [S/Mg] ratio at different metallicities). % and in any Galactic population.
This could reveal that there is no important variation in the yields of these two species with time, and
could again confirm their common origin in SNII explosion.
\end{itemize}
As a conclusion of the present subsection, it can be said that sulfur can be considered
 a good chemical clock, like any other $\alpha$-species, although some dispersion may be present
(see e.g. Santos-Peral et al. 2021, submitted).

\subsection{Sulfur in the different Galactic populations}
\label{Sect.Popul}
A dichotomy in the \alfaFe \  abundances, associated with the thin disc--thick disc bimodality,
has been found in the Galactic disc stellar populations
\cite[see e.g.][]{Vardan11, Recio14}. 
To date, such a chemical separation has been found either using an averaged  \alfaFe \ index (see above references), but also using different individual \alfa-species such as magnesium \citep[e.g. within the AMBRE context][]{Sarunas17, Pablo20} or other individual \alfaFe \ ratios \citep[][among others]{Sarunas14}. However, such a chemical dichotomy has never been established using only sulfur abundances. We note that in the very recent study of CS20 the thin--thick disc separation was first defined thanks to a global
\alfaFe \ abundance ratios and then applied to and/or checked against their sulfur content.
We decided to follow an opposite approach by looking for a possible definition of the Galactic components
based purely on their sulfur abundances. 
We indeed recall that, as already mentioned in Sect.~\ref{Sect.kinematics}, 
the dichotomy of the two disc components is clearly seen in kinematics and is correlated to the sulfur content.
This approach is also favoured since the thin--thick disc separation is expected
to slightly differ from one \alfa-species to another, as already shown by \cite{Sarunas14} or 
more recently by \cite{Amarsi20}.
Such different behaviours for different \alfa-elements could be real 
(slightly different nucleosynthesis channels) or caused by 
(among other possibilities) different internal dispersions resulting from different residual 
systematics such as non-LTE effects and/or different sensibilities of the selected lines between dwarfs and giants (the proportion  of dwarfs and giants in either disc being different due
to observational selection functions).

\begin{figure}
\resizebox{\hsize}{14cm}{\includegraphics{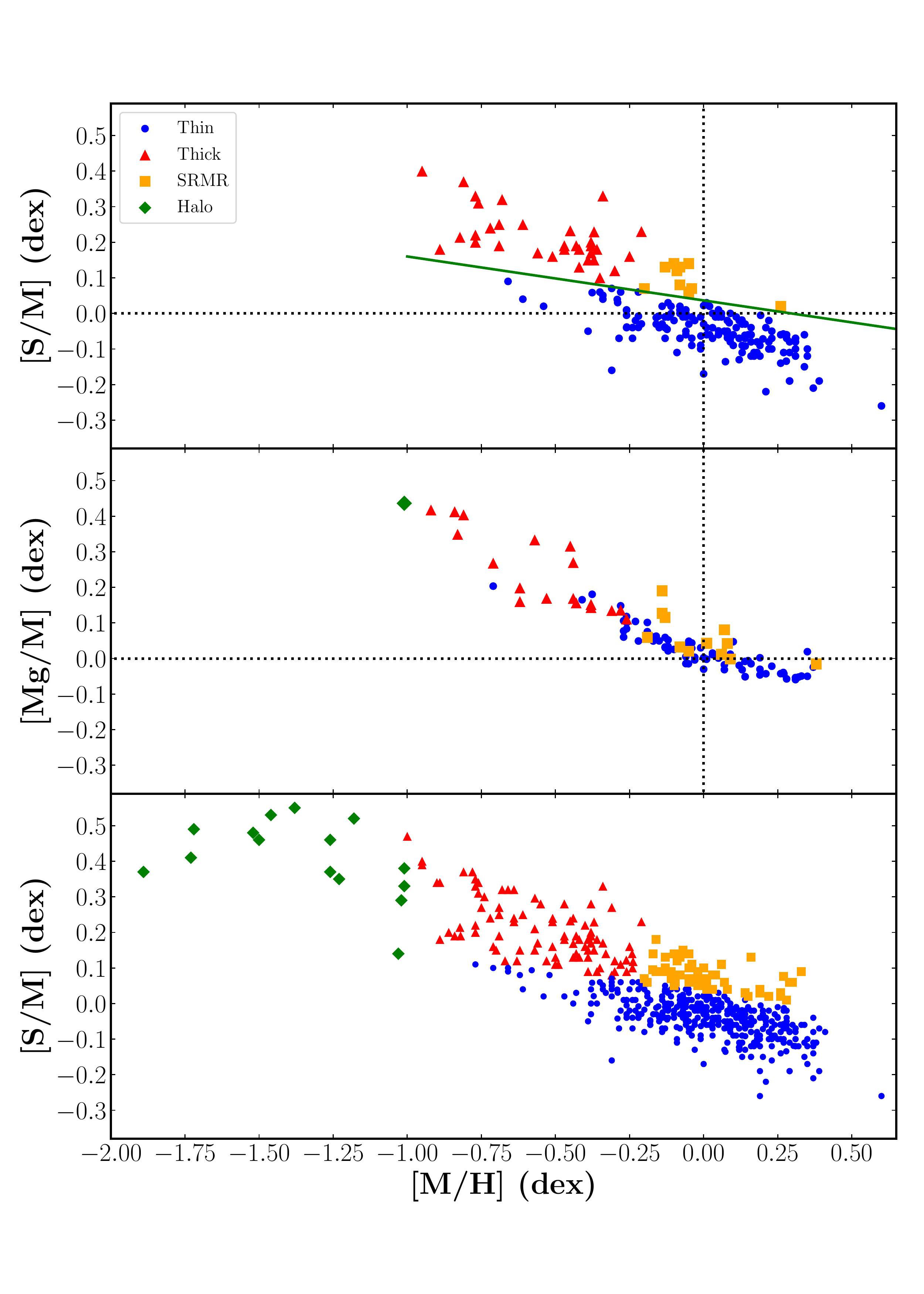}}
        \caption{$Top$: Ratio of sulfur abundance to mean metallicity   [S/M] vs.   mean 
metallicity \meta \ for stars with the best measured sulfur abundances (large number of analysed spectra, at least three measurements of the three S{\sc i} lines available
and small dispersion). The green line shows the adopted separation between the sulfur-rich and sulfur-poor
stars. $Middle$: [Mg/M] ratio vs. \meta \ for the stars in common with \cite{Pablo20}. $Bottom$: Same
as top panel, but for the complete sample of 540 stars with the best derived abundances (same sample as
in Fig. \ref{Fig.S_H-Best}, second panel), and labelled according to their Galactic population membership.
In all these panels thin-disc, thick-disc, sulfur-rich, metal-rich, and halo stars are shown as blue filled circles,
red triangles, orange squares, and green diamonds, respectively.}
\label{Fig.S-popul}
\end{figure}

Defining the sulfur-rich/sulfur-poor separation (associated hereafter with the \alfa-rich/\alfa-poor or 
thick-disc/thin-disc Galactic populations, respectively) could be useful, among other interests, to compute the radial chemical
gradients in both discs. For this purpose we first selected our best sulfur abundances by keeping only stars having a large number of analysed spectra.
More specifically, we considered 200 stars having more than 20 independent measurements of their 675.7~nm line,
more than 50 analysed lines in total including at least three measurements of
the three S{\sc i} lines, and, in addition, an internal
dispersion among all these measurements smaller than 0.05~dex. These best measured stars are shown in the top panel of Fig.~\ref{Fig.S-popul}.
We also defined a separation to disentangle the thin and  thick discs (green line in Fig.~\ref{Fig.S-popul}).
It was   drawn by examinating the [S/M] distributions in 0.2~dex wide metallicity bins for \meta \ varying within [-1.0,0.0]. Then, we looked for the position of the 
low-density regions separating both disc sequences, and drew a straight line
along these minima. This green line
was then   extrapolated towards  supersolar metallicites. Then, this separation was   
applied to both the entire and the {\it golden} samples.
The resulting $\sim$65\% sulfur-poor,  thin-disc stars
are shown as blue filled circles below the green line
in the bottom panel of Fig.~\ref{Fig.S-popul}. Moreover, for the sulfur-rich stars (found above the
separation line), a gap in their number distribution might be present around \meta=-0.2~dex.
Such a gap was already suspected among \alfa-rich stars around the same metallicity
by \cite{Vardan11}, who proposed to call them high-\alfa \ metal-rich stars \citep[see also][]{Gazzano13}. 
Although not clearly seen in sulfur and sometimes absent in other studies of \alfa \ elements,
depending on the analysed sample, we nevertheless decided to label  such stars separately.
This led to $\sim$18\% sulfur-rich metal-rich stars (hereafter  SRMR, and
shown as orange squares in Fig.~\ref{Fig.S-popul}) and $\sim$15\% thick-disc stars as red triangles. Finally, the stars more metal-poor than -1.0~dex were  labelled as   potential Galactic halo members (green diamonds). 

In confirmation of the above, several remarks can be made regarding the bottom panel of Fig.~\ref{Fig.S-popul}.  First,
it can be again seen that the sulfur abundances of   the thin-disc stars and the SRMR stars
continue to slowly decrease below [S/M]=0 when \meta \ becomes positive.
Then, most of the thin-disc stars have metallicity higher than -0.5~dex, although 
an extension can be seen down to
\meta$\sim$-0.9~dex. The thick-disc stars follow the slope of the green line, and are
preferentially located $\sim$0.1~dex above it. Finally, potential halo stars are found down to 
\meta$\sim$-2.0~dex. Their halo membership is confirmed by their rather low rotational velocities,
as shown in the top panel of Fig. \ref{Fig.kinematics}.
Their [S/M] ratio is compatible with a flat behaviour (like any other \alfa-element) and they have a mean [S/M] ratio close to +0.45~dex
with a dispersion equal to 0.15~dex. 

Moreover, we also show in the middle panel of Fig. \ref{Fig.S-popul} 
the stars with Mg 
abundances from \cite{Pablo20}, already shown in Fig. \ref{Fig.S-Mg}. They are drawn according to
their Galactic population membership as defined by their sulfur abundances. It can be seen
that (i) the thin-disc/sulfur-poor stars are the most magnesium poor, (ii) most of the SRMR stars
are also magnesium-rich/metal-rich, and (iii) thick-disc/sulfur-rich stars tend also 
to have higher [Mg/M] ratios. There is only one halo star in this subsample, but its sulfur
and magnesium enrichment are consistent. Therefore, although the scatter in sulfur appears larger than
in magnesium, we can conclude that defining the Galactic populations from their 
[S/M] content would give results consistent with those obtained from most commonly used ratios,  for instance [Mg/M].

\subsection{Galactic radial gradients in sulfur}
For the estimation of the Galactic gradient of sulfur for our $\sim$400 stars
belonging to the thin disc, we considered the guiding
centre radius ($R_{\rm g}$), computed as the average of the pericentre and apocentre
of the stellar orbits, as a proxy of the present star Galactocentric distance.
The gradient (and associated uncertainty) was   estimated thanks to a Theil-Sen fit of the \SM \ versus $R_{\rm g}$ points, the uncertainty being given by the lower and upper confidence levels of this fit.
We find a small positive radial gradient $\delta {\rm \SM} / \delta R_{\rm g}$~=~+0.004$^{\pm 0.006}$~dex/kpc in the thin disc 
for $R_{\rm g}$ within 6 and 10~kpc. This gradient is slightly flatter than the \alfaFe \ gradient
of 0.012$\pm$0.002~dex/kpc
found between 5 and 13~kpc using Gaia-ESO Survey abundances \cite[but no Gaia distances were
available at that time]{Recio14}
and that of \citet{Pablo20} estimated for magnesium over iron for AMBRE stars (+0.025$\pm$0.009~dex/kpc between 
6 et 11kpc). However, the Galactic (thin) disc gradient in sulfur has  recently been estimated from 17 H{\sc ii} 
regions with revisited Gaia distances ranging between $\sim$7 and $\sim$14~kpc\citep{HII}. 
This study reports a radial gradient in [S/H] equal to -0.035$^{\pm 0.006}$~dex/kpc in good agreement within the error bars with our determination: $\delta {\rm [S/H]} / \delta R_{\rm g}$~=
-0.05$^{\pm 0.025}$~dex/kpc. We also note that these authors found a flat gradient for [S/O], 
confirming independently   the similar nature of these two chemical species, and hence the $\alpha$  
nature of sulfur.

We  also computed the radial gradient in the thick disc between 4.5 and 15.5~kpc:
$\delta {\rm [S/M]} / \delta R_{\rm g}$~=~-0.014$^{\pm0.014}$~dex/kpc, in 
agreement within the error bars with the value reported by \citet[-0.004$\pm$0.003 dex within 4 and 11~kpc]{Recio14}.
Adding the SRMR stars to the thick-disc sample (leading to a total
of $\sim$150 stars) would not change this sulfur-to-mean metallicity radial gradient since we found -0.01$^{\pm-0.015}$~dex/kpc.

%\subsection{A Faire}
%Fig5: Regarder alpha des halo à meta=-1: Enceladus???? 
%Voir aussi les 4 thindiscs metalpoor et S/M~-0.1 qui se détachent du groupe

        \section{Summary}
        \label{Conclusions}
        We have presented LTE sulfur abundances derived from the three main components of the 
        multiplet~8 system lines found around 675~nm, which are known to be poorly affected by NLTE effects.
        This study analysed $\sim$100,000 spectra (including several repeats per stars)
        retrieved from the ESO archives of the HARPS, 
        FEROS, and UVES instruments. 
        These sulfur abundances have been homogeneously measured at a spectral resolution of 40,000 thanks to (i) stellar atmospheric parameters previously
        determined within the AMBRE Project \citep{AMBRE13}; 
        (ii) GAUGUIN, an optimisation method based on the Gauss-Newton algorithm;        and  (iii) a precomputed grid of synthetic spectra with [S/H]   abundances varying from -3.0
        to +2.0~dex.
        Then, each spectrum with at least one measured S{\sc i} line has been considered to derive a mean
        sulfur abundance per star and an associated dispersion, based on the line-to-line scatter and
        the available repeat spectra.   
        The final catalogue contains abundances for 1,855 individual slow-rotating stars. About 90\% of the
        sample consists of FGK-dwarf stars and the remaining 10\% are cool giants. Their mean metallicity
        is   between -2.0 to +0.7~dex.
        This is the largest catalogue of accurate and precise sulfur abundances published to date.
        We also present sulfur abundances of 13 $Gaia$ benchmark stars re-estimated by considering their recommended
        stellar parameters.

        This AMBRE-sulfur catalogue allowed us to study the origin and evolution of sulfur in the 
        Milky Way. First, we have shown that sulfur presents behaviours
        that are close to any other  $\alpha$-element. The mean [S/M]
        is equal to $\sim$0.45~dex for \meta < -1.0~dex and the distribution of [S/M] is 
        compatible with a plateau-like behaviour in the low-metallicity
        regime. Then, a monotonic decline is found with increasing metallicity. Moreover, this
        decline clearly continues for supersolar metallicity (without any slope change), 
        as already independently found for AMBRE magnesium abundances \citep{Pablo20}.
        All of this is also confirmed by the low [S/$\alpha$] ratios found over the whole studied metallicity range
        and by the similar behaviour of our sulfur abundances with metallicity compared to previous magnesium
        and europium abundances, both elements being predominantly produced by Type II supernovae (although
        still discussed for europium).
        
        Then, thanks to Gaia DR2 astrometry, we have studied the correlation between the AMBRE-sulfur
        abundances and the stellar kinematic and orbital properties. 
        A dichotomy in kinematics and
        eccentricity is detected between sulfur-rich and sulfur-poor stars at metallicities lower than $\sim$-0.5~dex. Two disc
        components that could be associated with the thin and the thick discs, with different sulfur content and kinematical
        properties are also identified. We have then proposed that the thin-disc/thick-disc dichotomy could be defined by solely 
        considering  
        sulfur abundances, as done in previous studies by considering any other \alfa-species. Furthermore, a trend with small dispersion between stellar ages and sulfur content
        is found: [S/M] slowly increases with stellar age up to $\sim$11~Gyr, whereas the metallicity decreases 
        and then a much steeper slope appears for older more metal-poor stars. Sulfur could thus be used as a chemical clock, although some dispersion could appear when examining a larger sample.
        Finally, we have estimated the sulfur radial gradient in the thin disc
        and found a small positive gradient for $\delta {\rm \SM} / \delta R_{\rm g}$ consistent with previous studies, 
        in particular the gradient derived from the sulfur content of H{\sc ii} regions.
        The gradient in the thick disc is found to be slightly smaller.

        This work  therefore proposes that the Galactic chemical history of sulfur is similar to 
        that of a 
        typical \alfa-element, and that this chemical species could be adopted with confidence to
        study   Galactic stellar populations.

        \begin{acknowledgements}
        We are grateful to Robin Bonannini who starts a preliminary
                study of sulfur abundances with AMBRE data several years ago.
        This work has made use of data from the European Space Agency (ESA) mission $Gaia$ (\href{https://www.cosmos.esa.int/gaia}{https://www.cosmos.esa.int/gaia}), processed by the $Gaia$ Data Processing and Analysis Consortium (DPAC, \href{https://www.cosmos.esa.int/web/gaia/dpac/consortium}{https://www.cosmos.esa.int/web/gaia/dpac/consortium}). Funding for the DPAC has been provided by national institutions, in particular the institutions participating in the $Gaia$ Multilateral Agreement. This research has also made use of the SIMBAD database, operated at CDS, Strasbourg, France. The authors acknowledge financial support from the ANR 14-CE33-014-01. This work was also supported by the "Programme National de Physique Stellaire" (PNPS) of CNRS/INSU co-funded by CEA and CNES. 
        M.A.A. acknowledges support from the CIGUS -CITIC, funded by Xunta de Galicia and the European Union (FEDER Galicia 2014-2020 Program) by grant ED431G 2019/01. 
        Finally, most of the calculations have been performed with the high-performance computing facility SIGAMM, hosted by OCA.
        \end{acknowledgements}
        
        % for the bibliography, at the end
        \bibliographystyle{aa} % style aa.bst
        \bibliography{ref} % your references Yourfile.bib
        
%       \begin{appendix}
%               \section{Title of the first appendix}
%               \section{Title of the second}
%       \end{appendix}
                
\end{document}